\documentclass[journal,twoside,web]{IEEEtran}
\usepackage{cite}
\usepackage{amsmath,amssymb,amsfonts}
\usepackage{graphicx}
\usepackage{hyperref}
\hypersetup{hidelinks=true}
\usepackage{textcomp}

\usepackage{graphics}
\usepackage{graphicx}
\graphicspath{{./fig/}}
\usepackage{color}
\usepackage{subfig}
\usepackage{amssymb}
\usepackage{amsmath}
\usepackage{ntheorem}
\usepackage{amsfonts}
\usepackage{mathrsfs}
\usepackage{latexsym, bm}
\usepackage{hyperref}
\usepackage{pdfpages}
\newtheorem{theorem}{Theorem}
\newtheorem{proposition}{Proposition}

\newtheorem{definition}{Definition}
\newtheorem{lemma}{Lemma}
\newtheorem{corollary}{Corollary}

\newtheorem{remark}{Remark}
\newtheorem{example}{Example}
\newtheorem{assumption}{Assumption}
\newtheorem*{problem}{Problem}
\newenvironment{proof}{\par\noindent\textbf{Proof.} }{\hfill$\blacksquare$\par}

\newcommand{\E}{\mathbb{E}}

\newcommand{\Var}{\textup{Var}}

\newcommand{\interior}{\operatorname{int}}
\newcommand{\norm}[1]{\| #1 \|}
\newcommand{\bd}[1]{\boldsymbol{#1}}
\newcommand{\Vf}{\mathcal{V}_{\textup{f}}}
\newcommand{\Vp}{\mathcal{V}_{\textup{p}}}
\newcommand{\Vn}{\mathcal{V}_{\textup{n}}}
 \DeclareMathOperator{\Span}{\mathbf{span}}

\newcommand{\dg}[1]{\bm{[}#1\bm{]}}

\usepackage{lineno}
\newcommand{\until}[1]{\{1,\dots, #1\}}

\newcommand{\setdef}[2]{\{#1 \; | \; #2\}}

\newcommand{\map}[3]{#1: #2 \rightarrow #3}

\newcommand{\V}{\mathcal{V}}
\newcommand{\MSE}{\textup{MSE}}
\newcommand\oprocendsymbol{\hbox{$\square$}}
\newcommand\oprocend{\relax\ifmmode\else\unskip\hfill\fi\oprocendsymbol}

\renewcommand{\theenumi}{(\roman{enumi})}

\DeclareSymbolFont{bbold}{U}{bbold}{m}{n}
\DeclareSymbolFontAlphabet{\mathbbold}{bbold}
\usepackage{algorithm}
\usepackage[noend]{algorithmic}
\algsetup{indent=2em}

\renewcommand{\algorithmiccomment}[1]{// \emph{#1}}
\DeclareMathOperator*{\argmin}{argmin}

\usepackage[prependcaption,colorinlistoftodos]{todonotes}

\begin{document}

%\journal{Systems and Control Letters}



\title{Susceptibility optimization for wisdom of crowds: A social power reformulation and realization framework}

\author{Ye Tian, Anzhi Sheng, Angela Fontan, Long Wang, and Karl H. Johansson \thanks{This work was partially supported by 
the Wallenberg AI, Autonomous Systems and Software Program (WASP) and the Wallenberg Scholar Grant funded by the Knut and Alice Wallenberg Foundation, the Swedish Research Council Distinguished Professor Grant, and the National Natural Science Foundation of
China (Grant Nos. 62533002 and 62036002).}
\thanks{Y. Tian, A. Sheng, A. Fontan and K. H. Johansson are with the Division of Decision and Control Systems, School of Electrical Engineering and Computer Science, KTH Royal Institute of Technology, and with Digital Futures, Stockholm 100 44, Sweden (e-mail: \{ye3; anzhi; angfon; kallej\}@kth.se).}
\thanks{L. Wang is with the Center for Systems and Control, College of Engineering, Peking University, Beijing 100871, China (e-mail: longwang@pku.edu.cn).}
}  %\ead{tinybeta7.1@gmail.com}

\maketitle
\begin{abstract}
Whether social influence makes a crowd wiser or more biased hinges on who yields to others and who holds firm. Such heterogeneous susceptibilities to social influence, together with the influence network, determine the weight  
each agent's initial estimate carries in the collective estimate, known as social power. This paper studies susceptibility optimization for maximizing the wisdom of crowds under the Friedkin--Johnsen model. 
Minimizing the squared collective bias over susceptibilities leads to a nonconvex problem. To overcome this nonconvexity, we develop a social power reformulation and realization framework, which decomposes the problem into two stages: first characterizing the social power that minimizes the collective bias, then determining the susceptibilities that induce such social power. Social power reformulation yields a quadratic program over the probability simplex. Under uniform-sign biases, the optimal social power concentrates on the least biased agents and immediately determines the optimal susceptibilities; under
mixed-sign biases, its geometry enables a linear-time construction algorithm.
For social power realization, we introduce the notions of susceptibility realizability and establish necessary and sufficient conditions for them, under which we derive all optimal susceptibilities in closed form using the Laplacian pseudoinverse and eigenvector centrality. These results yield a unified algorithm that solves the original problem globally, with worst-case complexity cubic in the number of agents when realizability holds. Finally, we develop a fully distributed implementation with linear communication complexity, illustrating how collective intelligence can emerge from local interactions. 
\end{abstract}
\section{Introduction}
\subsection{Background}
The question of how to effectively aggregate information initially dispersed among interacting agents is fundamental in collective intelligence~\cite{FAH:45,DC:22}. It arises naturally in networked systems where decisions rely on distributed and heterogeneous information, ranging from social networks and economic markets to multi-agent systems and, more recently, networks of LLM agents~\cite{doi:10.1073/pnas.1100290108,wang2024unleashing,yun2026graphofagents}.

A classical answer to this question is the wisdom of crowds effect, which suggests that simply pooling many independent estimates of an unknown truth yields collective estimates close to the truth~\cite{FG:1907}. Building on this principle, the Delphi method provides a structured and anonymous communication protocol for iterative aggregation of expert opinions, in which a moderator controls the information exchange process~\cite{ND-OH:63,OH:67}.  A key empirical finding from Delphi experiments is that those without strong convictions are more susceptible to social influence (\emph{swingers}), whereas those with strong arguments for deviant estimates tend to retain them (\emph{holdouts})~\cite{BG-BAM-JB:15,GM-GdP:15,JB-DB-DC:17}. These observations highlight that individuals exhibit heterogeneous susceptibilities to social influence, which are closely related to the quality of their private information.

Opinion dynamics provides a mathematical framework for studying how interpersonal interactions shape collective decisions~\cite{MHDG:74,NEF-ECJ:99,YT-LW:23}.
A growing body of work uses opinion dynamics models to explain how social influence affects the wisdom of crowds~\cite{BG-MOJ:10,FB-FF-BF:20,11312070,GC-FF-AP:25}. These studies suggest that, given individual accuracy, collective wisdom is completely determined by the social power allocation of the opinion dynamics~\cite{YT-LW-FB:23}, i.e., the relative control of each agent's initial opinion over the final collective decision. Among these models, the Friedkin--Johnsen (FJ) model~\cite{NEF-ECJ:99} is particularly relevant, as it incorporates heterogeneous susceptibilities that determine how much agents revise their opinions in response to social influence, and thus directly captures a spectrum of holdout--swinger behaviors, as supported by empirical evidence~\cite{NEF-ECJ:11}.

Studies of \emph{susceptibility to persuasion} suggest that susceptibility is not fixed but can be systematically shaped~\cite{mcguire1964inducing,ALLEN_PORTER_McFARLAND_2006}. In human-computer interaction,
persuasive technologies are designed to alter people's attitudes or behaviors~\cite{Fogg2003Persuasive}, illustrating that technological interventions can amplify or attenuate individuals' susceptibilities. From an engineering perspective, FJ-type models are well suited for modeling distributed decision tasks in networked systems, as susceptibility parameters explicitly quantify the tradeoff between private information and network interactions. Such models are increasingly relevant in modern engineered systems~\cite{MC-MY-LZ:26}, such as networks of LLM agents~\cite{yun2026graphofagents} and distributed optimization~\cite{11045805}, where agents interact through graph-structured protocols. Notably, recent work shows that the FJ model can capture the collective behavior of interacting LLM agents, with the social power acting as the gating weights of a mixture of experts~\cite{bause2026mixtures}. In these settings, susceptibilities are agent-specific parameters that allow both centralized tuning and decentralized adaptation. The FJ model has been used to study different interventions. In~\cite{10.1145/3466617}, susceptibilities are tuned to minimize the sum of final opinions, while in~\cite{CR-VB-PF-FD-MT:26}, incentives acting on agents' biases are designed to promote target actions. However, the problem of designing susceptibilities to maximize collective wisdom remains open.

\subsection{Contribution}
Motivated by the plasticity of susceptibility, this paper investigates susceptibility optimization for maximizing the wisdom of crowds. We formulate the problem under the FJ model, where $n$ agents form independent initial estimates of an unknown truth and update them under social influence. Our objective is to determine the optimal susceptibilities that minimize the squared collective bias, which yields a nonconvex optimization problem. To this end, we develop a \emph{social power reformulation and realization} (SPRR) framework. The key insight is that the collective estimate depends on susceptibilities exclusively through the social power. Accordingly, SPRR proceeds in two
stages: social power reformulation recasts the collective task over the social power simplex, whose solutions specify target social power; social power realization then seeks the susceptibilities that induce these targets, any of which is globally optimal for the original problem. In this way, SPRR reveals and exploits the hidden convexity of the
original problem in the social power space.

We first pose the social power realization problem for general influence systems.
Specializing to the FJ model, we define \emph{$\bd{x}$-susceptibility-realizability} ($\bd{x}$-SR) and \emph{full susceptibility-realizability} (FSR), which describe whether a prescribed social power can be induced by susceptibilities. We further establish the monotonicity and a linear characterization of the FJ social power, which interprets it as a redistribution of power and makes this hidden convexity explicit.

For the reformulation stage, we characterize the set of optimal social power under different bias structures. Under uniform-sign biases, the optimal social power concentrates on the least biased agents, immediately determining all optimal susceptibilities. When susceptibilities are bounded away from one, such social power is unattainable, and the problem coincides with the opinion minimization
problem~\cite{10.1145/3466617}. For this case, we identify an influence-bias hierarchy under which all optimal susceptibilities are explicitly characterized (Theorem~\ref{T0}), removing the need for local search and strong connectivity in~\cite{10.1145/3466617}.
Under mixed-sign biases, we show that the optimal solution set forms a convex polytope induced by a bipartite graph (Theorem~\ref{T1}). Based on this structure, we develop a linear-time algorithm that constructs an optimal social power vector by solving an edge-cover problem (Algorithm~\ref{alg:ospca}).

For the realization stage, we prove that the FJ model is FSR if and only if the influence network is strongly connected; under this condition, all susceptibilities realizing a given optimal social power are characterized in closed form via the \emph{Laplacian pseudoinverse} and \emph{eigenvector centrality} (Theorem~\ref{T3}). In particular,
the minimum collective bias remains attainable even when all agents are highly susceptible to social influence. More generally, we derive necessary and sufficient conditions for $\bd{x}$-SR, together with the corresponding optimal susceptibilities (Corollary~\ref{C0}). Combined with the optimal social power sets, these results yield all globally optimal susceptibilities, leading to an $\mathcal{O}(n^3)$ algorithm (Algorithm~\ref{alg:sspa}) when $\bd{x}$-SR holds. Otherwise, the algorithm still obtains a
globally optimal solution by solving a convex quadratic program (QP).

Finally, we show that both stages of SPRR admit fully distributed implementations under local information constraints: the edge cover underlying optimal social power construction is formed through one-hop pairing between agents with opposing biases, and the realization step
is carried out by linear iterations with guaranteed convergence (Theorem~\ref{T4}). This yields a distributed SPRR algorithm with linear communication complexity (Algorithm~\ref{alg:dsspa}).

Compared with our preliminary conference paper~\cite{YT-AS-AF-LW-KHJ:26}, this paper contains several substantial novel results: (i) the monotonicity of social power, a complete treatment of the uniform-sign bias case, and a social power construction algorithm for the mixed-sign bias case; (ii) a complete treatment of social power realization, including the notions of FSR and $\bd{x}$-SR with necessary and sufficient conditions for them, optimal susceptibilities under the $\bd{x}$-SR condition, and the resulting algorithm with complexity guarantees; and (iii) the decentralized implementation of SPRR.

\subsection{Outline}
Section~\ref{S2} formulates the problem. Section~\ref{S3} presents the SPRR framework, poses the social power realization problem, and establishes properties of social power. Sections~\ref{S4} and~\ref{S5} address optimal social power construction and realization, respectively. Section~\ref{S6} develops a decentralized implementation of SPRR, and Section~\ref{S7} concludes the paper.
 
\subsection{Notation}
Let $\bd{1}_{n}$, $\bd{e}_{i}$, and $I_{n}$ denote the all-ones vector, the $i$-th standard basis vector, and the $n\times n$ identity matrix, respectively. Given $\bd{x}\in\mathbb{R}^{n}$, $\dg{\bd{x}}$ denotes the diagonal matrix with diagonal entries $x_{1}, \dots, x_{n}$. For a square matrix $A$, $A^{\dagger}$ denotes its
Moore--Penrose pseudoinverse and $\rho(A)$ denotes its spectral radius. The $n$-simplex is defined as $\Delta_{n}=\setdef{\bd{x}\geq 0}{\bd{1}^{\top}_{n}\bd{x}=1}$, with interior $\interior{\Delta_{n}}$. A nonnegative matrix is row-stochastic (column-stochastic) if its row (column) sums equal $1$. 

The weighted digraph $\mathcal{G}(W)=(\mathcal{V},\mathcal{E})$ associated with a row-stochastic matrix $W$ consists of node set $\V=\until n$ and edge set $\mathcal{E}=\setdef{(i,j)}{W_{ij}>0}$, where $(i,j)$ is a directed edge from $i$ to $j$. Its Laplacian is $L=I_{n}-W$.
$\mathcal{G}(W)$ is strongly connected if there exists a directed path between every pair of distinct nodes; equivalently, $W$ and $L$ are irreducible. For an irreducible matrix $W$, the dominant left eigenvector $\bd{\omega}\in\interior\Delta_{n}$ satisfying $\bd{\omega}^{\top}W=\bd{\omega}^{\top}$ is unique and is called the eigenvector centrality of $\mathcal{G}(W)$.
A strongly connected component (SCC) of $\mathcal{G}(W)$ is called a sink SCC if it has no outgoing edges to others. 
An (unweighted) undirected bipartite graph $\mathcal{G}=(\mathcal{V}_{1},\mathcal{V}_{2},\mathcal{E})$ consists of two disjoint node sets $\mathcal{V}_{1}$ and $\mathcal{V}_{2}$, and a set of undirected edges $\mathcal{E}\subseteq\setdef{\{i,j\}}{i\in\mathcal{V}_{1},j\in\mathcal{V}_{2}}$. It is complete if $\mathcal{E}=\setdef{\{i,j\}}{i\in\mathcal{V}_{1},j\in\mathcal{V}_{2}}$.
 An edge cover $\mathcal{C}$ of $\mathcal{G}$ is a subset of
$\mathcal{E}$ such that every node is incident
to at least one edge in $\mathcal{C}$, and a minimum edge cover is one with the smallest cardinality.

\section{Problem formulation}\label{S2}

Consider $n$ agents collaboratively estimating an unknown state of nature $\theta\in\mathbb{R}$ over an influence network $\mathcal{G}(W)$. Due to localized information sources, specialized training datasets, or distinct prior experiences, agents exhibit heterogeneous expertise, which leads to heterogeneity in the quality of their private information.
Based on its private information, each agent $i$ independently forms an initial estimate $y_{i}(0)$, modeled as a random variable with mean $\mu_i\in\mathbb{R}$ and variance $\sigma_{i}^{2}>0$. 
Unlike the classical naive learning model where initial estimates are unbiased~\cite{BG-MOJ:10,FB-FF-BF:20,YT-LW-FB:23}, we assume $\mu_i\neq\theta, \; \forall i\in\V$. 
Let $\bd{\mu}=(\mu_1,\dots,\mu_n)^{\top}$ and $\bd{\delta}=\bd{\mu}-\theta\bd{1}_{n}$. Here, $\delta_i$ and
$\sigma_i^2$ capture the systematic bias and the uncertainty of agent $i$'s estimate, respectively.

Upon exposure to others' estimates, agents update their estimates according to the FJ opinion dynamics:
\begin{equation}\label{e1}
y_{i}(k+1)=a_{i}\sum_{j=1}^{n}W_{ij}y_{j}(k)+(1-a_{i})y_{i}(0),
\end{equation}
where $a_{i}$ is $i$'s susceptibility to social influence, and $1-a_{i}$ reflects its stubbornness to its own initial estimate. Let $\Vf$, $\Vp$, and $\Vn$ denote the sets of \emph{fully stubborn} ($a_i=0$), \emph{partially stubborn} ($0<a_i<1$), and \emph{nonstubborn} ($a_i=1$) agents, respectively.
Since $W_{ii}=1$ is equivalent to $a_i=0$, to avoid triviality, we assume $W_{ii}\neq 1, \; \forall i\in\V$. Let $\bd{a}=(a_{1},\dots,a_{n})^{\top}$ and $\bd{y}(k)=(y_1(k),\dots,y_n(k))^{\top}$. The following assumption ensures that $\bd{y}(k)$ has a closed-form limit.

\begin{assumption}\label{A1}
There is no sink SCC in $\mathcal{G}(W)$ that consists only of nonstubborn agents. 
\end{assumption}

\begin{lemma}\label{L0}\cite[Lemma 4]{YT-LW:23}
For the FJ model~\eqref{e1}, $\rho(\dg{\bd{a}}W)<1$ if and only if Assumption~\ref{A1} holds. In this case, 
\begin{equation}
\bd{y}(\infty)=\lim_{k\to\infty}\bd{y}(k)=(I_{n}-\dg{\bd{a}}W)^{-1}(I_{n}-\dg{\bd{a}})\bd{y}(0).
\end{equation} 
\end{lemma}

We denote by $\mathcal{D}_W=\setdef{\bd{a}\in[0,1]^n}{\rho(\dg{\bd{a}}W)<1}$ the set of susceptibility vectors satisfying Assumption~\ref{A1}. Let $V=(I_{n}-\dg{\bd{a}}W)^{-1}(I_{n}-\dg{\bd{a}})$ and define $y_{\textup{ave}}=(1/n)\sum_{i=1}^{n}y_{i}(\infty)$ as the collective estimate. Then, we have 
\begin{align}\label{equ:ce}
  y_{\textup{ave}}=\frac{1}{n}\bd{1}^{\top}_{n}V\bd{y}(0)=\bd{x}^{\top}\bd{y}(0),
\end{align}
where 
\begin{align}\label{equ:sp}
\bd{x}=V^{\top}\frac{\bd{1}_{n}}{n}=(I_{n}-\dg{\bd{a}})(I_{n}-W^{\top}\dg{\bd{a}})^{-1}\frac{\bd{1}_{n}}{n}
\end{align}
is the social power of the FJ model~\cite{YT-LW:23}, measuring the dominance of agents' initial estimates on the collective outcome. 
Since $\{y_i(0)\}_{i=1}^n$ are independent, we have 
\begin{align}\label{eqmv}
  \E [y_{\textup{ave}}] =\sum_{i=1}^{n}x_{i}\mu_{i}, \quad \textup{and} \quad \Var[y_{\textup{ave}}]=\sum_{i=1}^{n}x_{i}^{2}\sigma_{i}^{2}.
\end{align}

By~\eqref{eqmv}, the collective accuracy is jointly 
determined by the influence matrix $W$ and the susceptibility vector $\bd{a}$. 
In practice, intervening on susceptibilities is more feasible than redesigning the influence network. In engineered systems, 
$\mathcal{G}(W)$ is typically constrained by physical or hardware architectures, such as communication topologies in sensor networks or wiring structures in power grids, whereas $\bd{a}$ is simply a set of local programmable parameters. 
In social networks, reshaping $\mathcal{G}(W)$ faces even more pronounced structural and cognitive inertia, as interpersonal influence depends not only on exposure but also on subjective appraisal~\cite{MHDG:74}. An individual may be fully aware of their neighbors' opinions yet assign zero influence weight to them, as in bounded confidence models~\cite{VDB-JMH-JNT:09}. Meanwhile, severing an established tie is particularly challenging~\cite{doi:10.1126/science.1116869}. By contrast, $\bd{a}$ can be modulated more flexibly through external interventions or self-adaptation~\cite{mcguire1964inducing,Fogg2003Persuasive}, such as anti-fraud education or self-reflection. 
In terms of complexity, redesigning $\mathcal{G}(W)$ involves up to $n^2$ pairwise parameters, whereas $\bd{a}$ consists of only $n$ local parameters. This motivates the following problem. 

\begin{problem}(Susceptibility optimization)
Given the influence network $\mathcal{G}(W)$ and the bias vector $\bd{\delta}$, design the susceptibility vector $\bd{a}$ that solves
\begin{equation}\label{eq3}
 \min_{\bd{a}\in \mathcal{D}_W} f(\bd{a}),
\end{equation}
where $f(\bd{a})=(\E[y_{\textup{ave}}]-\theta)^{2}$ is the squared collective bias.
\end{problem}

Throughout the paper, we assume $\bd{\delta}\notin \Span\{\bd{1}_{n}\}$, since otherwise $f$ does not depend on $\bd{a}$. 

\begin{remark}(Measures of collective wisdom)
The overall accuracy is commonly quantified by the \emph{mean square error} (MSE):
\begin{equation}\label{MSE}
\begin{aligned}
    \MSE[y_{\textup{ave}}]
    =\E[(y_{\textup{ave}}-\theta)^2]=f(\bd{a})+\Var[y_{\textup{ave}}].
\end{aligned}
\end{equation}
Most theoretical studies assume that agents' independent estimates are unbiased~\cite{YT-LW:23}, under which $f(\bd{a})\equiv 0$. Therefore, collective wisdom is evaluated solely through $\Var[y_{\textup{ave}}]$. Empirical studies~\cite{JL-HR-FS-DH:11,JB-DB-DC:17}, however, suggest that individuals' estimates are typically biased, so that $f(\bd{a})\neq 0$ in general. In this case,
minimizing the MSE directly may trade a larger bias for a reduction in variance, yet a small variance around a biased mean still yields poor collective performance. Hence, a lexicographic formulation is more relevant:
\begin{equation}\label{eqMSE}
\begin{aligned}
  \min_{\bd{a}\in\mathcal{D}_W} \quad &\Var[y_{\textup{ave}}]\\
  \textup{s.t.} \quad\quad &\bd{a}\in\argmin_{\bd{a}^{\prime}\in \mathcal{D}_W} f(\bd{a}^{\prime}),
\end{aligned}
\end{equation}
whose first step is exactly problem~\eqref{eq3}. The SPRR framework readily extends to~\eqref{MSE} and~\eqref{eqMSE}; see Remark~\ref{R:variance}.
\end{remark}

In the FJ model, susceptibilities capture the tension between private information and social influence, as observed in group experiments~\cite{NEF-ECJ:11}. Hence, problem~\eqref{eq3} provides a principled framework for harnessing knowledge dispersed among agents. Note that problem~\eqref{eq3} assumes $\bd{\delta}$ is known. In practice, $\bd{\delta}$ can be either measured on benchmark tasks with known ground truth in the same domain, or learned by neighboring agents from their shared track record on resolved tasks; see Remark~\ref{Rde}. Beyond social networks, this framework also extends to engineered systems, as outlined in the following remark. 

\begin{remark}(Application scenarios)
 Problem~\eqref{eq3} arises in a range of collective estimation settings.
  (i) \emph{Human expert networks}: in the Delphi method, problem~\eqref{eq3} indicates how the moderator should intervene on participants' susceptibilities based on their domain biases, especially when $\mathcal{G}(W)$ is fixed, e.g., in roundtable discussions or institutional committees.
  (ii) \emph{AI-agent networks}: for AI agents trained on heterogeneous datasets or specialized in different domains, biases arise from training data, while susceptibilities are programmable, making problem~\eqref{eq3} directly implementable. 
  (iii) \emph{Sensor and robotic networks}: in cooperative
localization and navigation, local measurements are subject to systematic bias due to hardware imperfections or deployment conditions, so susceptibilities can be tuned via
problem~\eqref{eq3} to balance the reliance on local measurements against that on neighbors' estimates.
\end{remark}

\begin{figure}[!t]
\begin{center}
\includegraphics[width=0.85\hsize]{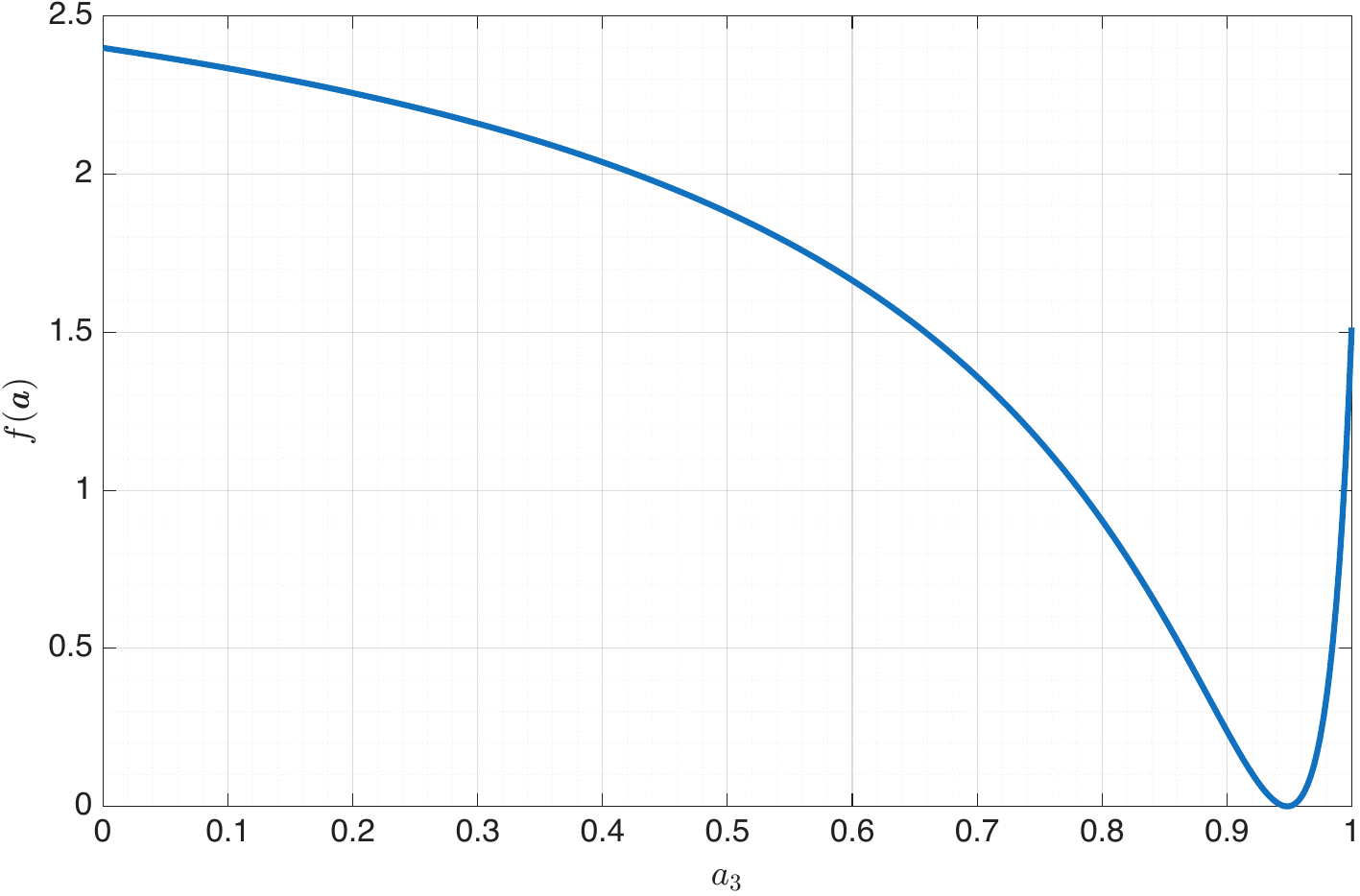}   
\caption{Example~\ref{exa1}. The squared collective bias $f(\bd{a})$ as a function of $a_3$, with $a_1$ and $a_2$ fixed, indicating that the considered problem is nonlinear and nonconvex.} 
\label{Fig1}
\end{center}
\end{figure}

In~\eqref{eq3}, the objective $f(\bd{a})$ is nonlinear and neither convex nor concave, as shown in the following example.  
\begin{example}(Nonconvexity of the objective)\label{exa1}
Consider an influence network $\mathcal{G}(W)$ with $n=3$, $\theta=0$, 
\begin{align*}
W=\begin{bmatrix}
   0.3 & 0 & 0.7 \\
0.2 & 0.3 & 0.5 \\
0.5 & 0 & 0.5
\end{bmatrix}, \quad \textup{and} \quad
\bd{\mu}=\begin{bmatrix}
-2\\
1\\
2
\end{bmatrix}.
\end{align*}
Fixing $a_1=0.9$ and $a_2=0.3$, we vary $a_{3}\in [0,1]$. Fig.~\ref{Fig1} shows $f(\bd{a})$ as a function of $a_3$. The curve is evidently neither convex nor concave.
\end{example}

Since $f(\bd{a})$ is differentiable, a natural approach to
problem~\eqref{eq3} is projected gradient descent. However, as $f(\bd{a})$ is nonconvex in $\bd{a}$, standard convex optimization guarantees do not apply directly. Moreover, evaluating the gradient (see Lemma~\ref{L5}) requires the global quantities $\bd{x}$ and $V$ with repeated matrix inversions, making gradient-based methods computationally demanding. In addition, such methods yield at most one optimal solution, without revealing the structure of the optimal solution set.

\section{Social power reformulation and realization}\label{S3}
\begin{figure}[!t]
\begin{center}
\includegraphics[width=\hsize]{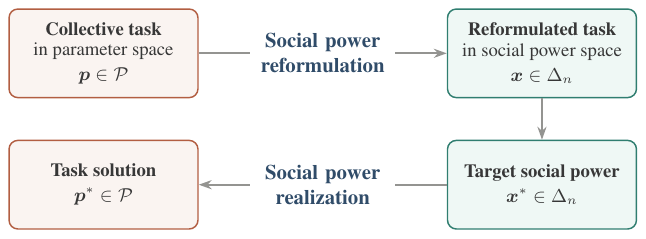}   
\caption{Diagram of SPRR. For an influence system with admissible parameter space $\mathcal{P}$, SPRR first reformulates a collective task defined over $\mathcal{P}$ as a task over the social power simplex $\Delta_n$. Solving the reformulated task yields a target
social power $\bd{x}^{*}\in\Delta_n$, whose realization gives a task solution $\bd{p}^{*}\in\mathcal{P}$. For problem~\eqref{eq3}, $\bd{p}=\bd{a}$ and $\mathcal{P}=\mathcal{D}_W$.}
\label{Fig12}
\end{center}
\end{figure}

This section introduces the SPRR framework and the social power realization problem, and establishes structural properties of social power that support the subsequent analysis. 
By \eqref{eqmv}, $f(\bd{a})$ depends on $\bd{a}$ only through the induced social power $\bd{x}$. 
Therefore, instead of optimizing over $\bd{a}$ directly, SPRR proceeds in two stages: it
first determines the optimal social power $\bd{x}^*$ that minimizes the squared collective bias, and then seeks susceptibility vectors $\bd{a}^*$ that realize $\bd{x}^*$. This two-stage framework is illustrated in Fig.~\ref{Fig12}.

In the first stage of SPRR, social power reformulation, we rewrite the squared collective bias in terms of social power $\bd{x}$. Define $\map{g}{\Delta_n}{\mathbb{R}}$ by $g(\bd{x})=(\bd{x}^{\top}\bd{\delta})^2$. Then by~\eqref{eqmv}, $g(\bd{x})=f(\bd{a})$ whenever $\bd{x}$ is the social power induced by $\bd{a}$. We consider optimizing the squared collective bias over $\bd{x}$:
\begin{equation}\label{eq11}
    \min_{\bd{x}\in\Delta_{n}} g(\bd{x}),
\end{equation}
which is a convex QP over the simplex, since $\bd{\delta}\bd{\delta}^{\top}$ is positive semidefinite. We characterize
the optimal social power $\bd{x}^*$ in Section~\ref{S4}. 
The second stage of SPRR, social power realization, seeks
susceptibility vectors that induce $\bd{x}^*$, any of which is globally optimal for problem~\eqref{eq3}. This leads to the following social power realization problem.

\subsection{The Social Power Realization Problem}

As shown in~\eqref{equ:ce}, the final collective estimate of the FJ model is given by $y_{\textup{ave}}=\bd{x}^{\top}\bd{y}(0)$. This aggregation form applies to any weighted-average influence system~\cite{YT-LW-FB:23}. A social power vector thus specifies a mapping from individual initial opinions to the final collective opinion. 
Achieving a prescribed aggregation behavior therefore requires selecting system parameters that induce the associated social power. Consider such an influence system with an admissible parameter space $\mathcal{P}$. A natural question is whether a prescribed social power $\bd{x}\in\Delta_n$ can be induced by some parameters $\bd{p}\in\mathcal{P}$. Analogous to the classical
state-space realization problem~\cite{astrom2021feedback}, which seeks a state-space model reproducing a prescribed input-output behavior, we refer to this problem as social power realization. 

Different choices of $\mathcal{P}$ yield different social
power realization problems. In the FJ model with fixed susceptibilities, taking $\mathcal{P}$ as a set of row-stochastic matrices leads
to influence network realization of social power. In particular, when $\bd{a}=a\bd{1}_n$, the FJ social power coincides with the PageRank vector of $W$ with damping factor $a\in (0,1)$, and enforcing a prescribed PageRank vector by perturbing edge weights~\cite{SC-FD-BM:25} is an instance of this problem. For problem~\eqref{eq3}, the influence matrix $W$ is fixed, $\mathcal{P}=\mathcal{D}_W$, and $\bd{p}=\bd{a}$. The social power realization problem therefore amounts to determining whether there exist susceptibility vectors that induce a target social power $\bd{x}$ and, if so, characterizing all such vectors. We introduce the following notions of susceptibility
realizability.

\begin{definition}\label{D1}(Notions of susceptibility realizability)
 The FJ model is said to be 
 \begin{enumerate}
  \item $\bd{x}$-susceptibility-realizable ($\bd{x}$-SR) if for a given target social power $\bd{x}\in\Delta_{n}$ there exists a susceptibility vector $\bd{a}\in \mathcal{D}_W$ that induces $\bd{x}$;
  \item fully susceptibility-realizable (FSR) if it is $\bd{x}$-SR for all $\bd{x}\in\Delta_{n}$.
 \end{enumerate}
\end{definition}

These notions characterize the extent to which the susceptibility vector can be used as a tuning knob to steer the social power, reflecting an intrinsic property of the FJ model. They are therefore relevant to any collective task governed by social power, as well as to FJ variants with a well-defined mapping from susceptibilities to social power~\cite{YT-LW:23,10849573}. Necessary and sufficient conditions for them are established in Section~\ref{S5}.

\subsection{Structural properties of social power}

We end this section with structural properties of the FJ social power $\bd{x}$, whose proofs are provided in Appendix~\ref{app:L}. 
In $\mathcal{G}(W)$, a directed path is said to be a \emph{partially stubborn path} (PSP) if all its intermediate nodes are partially stubborn. The following lemma follows from arguments similar to those of~\cite[Lemma 5]{YT-LW:23} and~\cite[Lemma A.3]{YT-YK-WZ-KK:25}.
\begin{lemma}\label{LV}
Suppose that Assumption~\ref{A1} holds. For
$V=(I_{n}-\dg{\bd{a}}W)^{-1}(I_{n}-\dg{\bd{a}})$, the following
statements hold:
\begin{enumerate}
  \item \label{LV-1} $V$ is row-stochastic, with $V_{ij}=0$ for all
  $i\in\V$ and $j\in\Vn$, and $V_{ii}>0$ for all $i\in\Vf\cup\Vp$;
  \item \label{LV-5} $\bd{x}\in\Delta_n$, and for all $i\in\V$, $x_i=0$
  if and only if $a_i=1$.
\end{enumerate}
If, in addition, $\bd{a}\in[0,1)^n$, then
\begin{enumerate}
  \setcounter{enumi}{2}
  \item \label{LV-3} $V_{ii}>V_{ji}$ for all $i\in\V$ and $j\neq i$;
  \item \label{LV-4} for all $i\in\Vp$ and $j\neq i$, $V_{ij}>0$ if and
  only if there exists a PSP from $i$ to $j$ in $\mathcal{G}(W)$.
\end{enumerate}
\end{lemma}

We now prove that $\bd{x}$ is elementwise monotonic in $\bd{a}$. Since $\bd{x}$ depends continuously on $\bd{a}$, it suffices to focus on $\bd{a}\in(0,1)^n$, under which Assumption~\ref{A1} holds.

\begin{proposition}(Monotonicity of social power)\label{Pmono}
Suppose that $\bd{a}\in (0,1)^{n}$. Then, 
\begin{enumerate}
    \item $x_{i}$ strictly decreases in $a_{i}$ for all $i\in\V$; \label{L1-1}
    \item for all $i\in\V$ and $j\neq i$, $x_{j}$ is nondecreasing in $a_{i}$; it strictly increases in $a_{i}$ if and only if there exists a PSP from $i$ to $j$ in $\mathcal{G}(W)$. \label{L1-2}
\end{enumerate}
\end{proposition}

Proposition~\ref{Pmono} shows that raising an agent's susceptibility transfers part of its social power to the agents reachable from it via PSPs, leaving all others unaffected.
Moreover, its proof yields the gradient of $f$ with respect to $\bd{a}$.

\begin{lemma}(Gradient of the squared collective bias)\label{L5}
Suppose that $\bd{a}\in (0,1)^{n}$. Then,
\begin{align}
\frac{\partial f}{\partial \bd{a}}=-2(\bd{x}^{\top}\bd{\delta})\dg{\bd{x}}(I_{n}-\dg{\bd{a}})^{-1}(I_{n}-WV)\bd{\delta}.
\end{align}
\end{lemma}

By~\eqref{equ:sp}, $\bd{x}$ depends nonlinearly on $\bd{a}$ through a matrix inverse in which $\bd{a}$ and $W$ intertwine. The next proposition shows that $\bd{x}$ admits a linear characterization that decouples $\bd{a}$ from $W$.

\begin{proposition}(Linear characterization of social power)\label{P4}
For the FJ model and $\bd{x}\in\mathbb{R}^{n}$, Assumption~\ref{A1} holds and $\bd{x}$ is its social power if and only if there exists $\bd{\gamma}\geq 0$ such that
\begin{subequations}\label{eq12}
\begin{align}
  \bd{\gamma}&=\dg{\bd{a}}(\bd{x}+\bd{\gamma}), \quad \textup{and} \label{eq12a}\\
  L^{\top}\bd{\gamma}&=\frac{\bd{1}_{n}}{n}-\bd{x},\label{eq12b}
\end{align}
\end{subequations}
where $L=I_n-W$ is the Laplacian of $\mathcal{G}(W)$.
\end{proposition}

Unlike the spectral characterization in our preliminary
work~\cite{YT-AS-AF-LW-KHJ:26}, which requires
$\bd{a}\in[0,1)^n$, Proposition~\ref{P4} covers nonstubborn agents, and thus, by Lemma~\ref{LV}~\ref{LV-5}, social power on the boundary of $\Delta_n$. Combining~\eqref{eq11} with Proposition~\ref{P4}, the two stages of SPRR can be cast into a single convex QP equivalent to
problem~\eqref{eq3}: $\min_{\bd{x}\in\Delta_n,\,\bd{\gamma}\geq 0}g(\bd{x})$ subject to~\eqref{eq12b}, with $\bd{a}$ determined by~\eqref{eq12a}. Hence, by reformulating problem~\eqref{eq3} in terms of social power, SPRR reveals its hidden convexity. However, solving this convex program numerically yields only one optimal solution. In the following sections, we characterize the optimal social power and all susceptibilities realizing it in closed form.

\begin{remark}(A redistribution perspective on social power)\label{R:redis}
By Proposition~\ref{P4}, rather than the accumulated influence of initial opinions, social power can be viewed as the outcome of a redistribution of power: $y_{\textup{ave}}$ endows each agent $i$ with equal power $1/n$; agent $i$ delegates power $W_{ij}\gamma_i$ to, and receives $W_{ji}\gamma_j$ from, each agent $j$. Thus, agent $i$ has incoming power $1/n+\sum_{j\in\V}W_{ji}\gamma_j$ and delegates $\gamma_i$ in total. By~\eqref{eq12b}, it retains $x_i$; by~\eqref{eq12a}, it delegates a fraction $a_i$ of its incoming power and retains the fraction $1-a_i$, which means that susceptibility determines how much power each agent delegates. Proposition~\ref{P4} also admits a random-walk interpretation: in the walk that traces back the source of a uniformly chosen agent's final opinion, $x_i$ and $\gamma_i$ are the probability of stopping at and the expected number of exits from agent $i$, respectively, as in stopping rules for Markov chains~\cite{BL10}.
\end{remark}

\section{Optimal social power construction}\label{S4}

This section characterizes the set $\mathcal{X}$ of optimal social power for problem~\eqref{eq11}. The geometry of $\mathcal{X}$ depends on the sign pattern of the bias vector
$\bd{\delta}$. We study the problem under two bias regimes: uniform-sign biases and mixed-sign biases.  

\subsection{Uniform-sign biases}
 
Under uniform-sign biases, all agents either overestimate or underestimate the truth, i.e., $\bd{\delta}> 0$ or $\bd{\delta}< 0$. Since $(\bd{x}^{\top}\bd{\delta})^2=(\bd{x}^{\top}(-\bd{\delta}))^2$, we assume $\bd{\delta}> 0$ without loss of generality, and let $\mathcal{I}_{\min}^{\bd{\delta}}=\argmin_{j}\delta_j$. Then, problem~\eqref{eq11} is solved by 
\begin{align}\label{eqosp}
  \bd{x}^{*}\in\mathcal{X}=\setdef{\bd{x}\in\Delta_{n}}{\textstyle\sum_{i\in \mathcal{I}_{\min}^{\bd{\delta}}}x_{i}=1}, 
\end{align}
which means optimal social power concentrates entirely on agents in $\mathcal{I}_{\min}^{\bd{\delta}}$. Since $\bd{x}^{*}\in\mathcal{X}$ lies on the boundary of
$\Delta_n$, its realization can be characterized directly. By Assumption~\ref{A1} and Lemma~\ref{LV}~\ref{LV-5}, the FJ model is $\bd{x}^{*}$-SR for some
$\bd{x}^{*}\in\mathcal{X}$ if and only if every sink SCC of
$\mathcal{G}(W)$ contains an agent in
$\mathcal{I}^{\bd{\delta}}_{\min}$. In this case, $\bd{a}^{*}$ solves problem~\eqref{eq3} if and only if 
\begin{align}\label{opti0}
  a^{*}_{i}=1, \quad \forall  i\notin\mathcal{I}_{\min}^{\bd{\delta}},
\end{align}
and each sink SCC contains an agent $j\in\mathcal{I}_{\min}^{\bd{\delta}}$ with $a^{*}_{j}<1$. 
This result highlights that, under uniform-sign biases, the optimal strategy is to make the least biased agents the anchors while making all others nonstubborn, consistent with experiments~\cite{JB-DB-DC:17}. The case where some sink SCC contains no agent in $\mathcal{I}_{\min}^{\bd{\delta}}$ is addressed in Section~\ref{S5}. 

Moreover, under the additional constraint
$\bd{a}\leq\bd{u}<\bd{1}_n$, no $\bd{x}^{*}\in\mathcal{X}$ can be induced. Note that, for $\bd{\delta}> 0$, minimizing $f(\bd{a})$ is equivalent to minimizing $\bd{1}_{n}^{\top}V \bd{\delta}$. Thus, problem~\eqref{eq3} coincides with the opinion minimization problem in~\cite{10.1145/3466617} with $\bd{y}(0)=\bd{\delta}$. For strongly connected $\mathcal{G}(W)$ and box constraints $0<l_i\leq a_{i}\leq u_i<1$, it is shown in~\cite{10.1145/3466617} that the sign of $\partial (\bd{1}_{n}^{\top}V\bd{\delta})/\partial a_{i}$ is independent of $a_{i}$, so that the minimum is attained at either $a_i=l_i$ or $a_i=u_i$. However, this sign is not explicitly determined, and local search algorithms are used to find the optimum. 
We next show that, for any $\mathcal{G}(W)$, least biased agents directly influenced by more biased agents are fully stubborn in every optimal solution.
Then, we establish an influence-bias hierarchy that determines the sign of each $\partial f/\partial a_i$. This explicitly characterizes all optimal solutions without requiring local search and strong connectivity.

\begin{theorem}\label{T0}
Suppose that $\bd{\delta}> 0$. Let $\mathcal{I}_{\min}^{\bd{\delta}}=\argmin_{j}\delta_j$, $\mathcal{J}=\setdef{i\in\mathcal{I}_{\min}^{\bd{\delta}}}{\textstyle
\sum_{j\notin\mathcal{I}_{\min}^{\bd{\delta}}}W_{ij}>0}$, $0<\bd{u}<\bd{1}_n$, and $\bd{\xi}=L\bd{\delta}$. Then, for problem~\eqref{eq3} subject to $\bd{a}\leq\bd{u}$,
\begin{enumerate}
  \item \label{P3-1}every optimal solution satisfies $a_i^*=0$ for all $i\in\mathcal{J}$;
  \item \label{P3-2}if 
  \begin{align}\label{equ12}
    \xi_i> 0, \quad \forall i\notin\mathcal{I}_{\min}^{\bd{\delta}},
  \end{align}
  then $\bd{a}^*$ is an optimal solution if and only if
\begin{align}\label{opti}
 a^{*}_{i}\begin{cases}
  =0 & i\in\mathcal{J},\\
  \in [0,u_i] & i\in\mathcal{I}_{\min}^{\bd{\delta}}\setminus\mathcal{J},\\
  =u_i & i\notin\mathcal{I}_{\min}^{\bd{\delta}}.
  \end{cases}
\end{align}
\end{enumerate}
\end{theorem}

The proof is provided in Appendix~\ref{app2}. Condition~\eqref{equ12} defines an influence-bias hierarchy on the influence network: for every agent whose bias is not minimal, the weighted average bias of its neighbors is smaller than its own. Hence, social influence tends to flow from more accurate agents to less accurate ones. Even if this hierarchy holds only non-strictly, maximal wisdom is achieved by making the least biased agents fully stubborn, while all others take their maximum admissible susceptibilities, as shown
in the proof. Moreover, since $\mathcal{I}^{\bd{\delta}}_{\min} = \argmin_j \mu_j$ and $\bd{\xi} = L\bd{\mu}$, the optimal susceptibilities in~\eqref{opti0} and~\eqref{opti}, as well as the hierarchy~\eqref{equ12}, can be determined from the agents' means $\bd{\mu}$ alone, without knowing the truth $\theta$.

\subsection{Mixed-sign biases}
Next, we study the case of mixed-sign biases.

\begin{assumption}\label{A3}
$\min_{i}\delta_{i}<0<\max_{i}\delta_{i}$. 
\end{assumption}

Under Assumption~\ref{A3}, $\mathcal{X}$ admits a complete geometric characterization: it is a convex polytope whose vertices are in one-to-one correspondence with the edges of a complete bipartite graph associated with the signs of the biases. This leads to a linear-time optimal social power construction algorithm. 

\begin{theorem}\label{T1}
Suppose that Assumption~\ref{A3} holds. Then, 
\begin{enumerate}
  \item the optimal solution set of problem~\eqref{eq11} is $\mathcal{X}=\setdef{\bd{x}\in\Delta_n}{\bd{x}^{\top}\bd{\delta}=0}$; \label{T1-1}
  \item Algorithm~\ref{alg:ospca} computes an optimal social power $\bd{x}^{*}\in\interior\Delta_n$ in $\mathcal{O}(n)$ time. \label{T1-2}
\end{enumerate}
\end{theorem}
\begin{algorithm}
\caption{Optimal Social Power Construction Algorithm (OSPCA)}
\label{alg:ospca}
\begin{algorithmic}[1]
  \REQUIRE  Bias vector $\bd{\delta}$ 
\ENSURE Optimal social power $\bd{x}^{*}\in\interior\Delta_n$
\STATE Partition $\mathcal{V}$ into $\mathcal{V}^{+}=\{i \in \mathcal{V} \mid \delta_i > 0\}$ and $\mathcal{V}^{-}=\{i \in \mathcal{V}\mid \delta_i < 0\}$
\STATE Compute an edge cover $\mathcal{C} \subseteq \mathcal{E}$ of the undirected complete bipartite graph $\mathcal{G}=(\mathcal{V}^{+},\mathcal{V}^{-},\mathcal{E})$ 
\STATE For each $i\in\V$, choose a $q(i)$ such that $\{i,q(i)\}\in\mathcal{C}$ 
and compute a mixing weight $\eta_i \gets \frac{\delta_{q(i)}}{(\delta_{q(i)} - \delta_i)}$
\STATE Pick $\bd{z} \in \interior \Delta_n$ 
\STATE For each $i\in\V$,  $x_{i}^{*}\gets z_{i}\eta_i+\sum_{l\in\setdef{j\in\V}{q(j)=i}}z_{l}(1-\eta_l)$
\RETURN $\bd{x}^{*}$
\end{algorithmic}
\end{algorithm}

Theorem~\ref{T1}, proved in Appendix~\ref{app2}, shows that, unlike the uniform-sign case, $\mathcal{X}$ always contains positive optimal social power, under which every agent contributes to the collective estimate and, by
Lemma~\ref{LV}~\ref{LV-5}, no agent needs to be nonstubborn. Such an $\bd{x}^{*}\in\interior \Delta_n$ can be constructed in linear time. The key idea is to achieve global bias cancellation by constructing pairwise mixing weights between agents with opposite-sign biases such that each local pairing cancels out its bias contribution. Aggregating all such mixing weights yields a social power allocation that eliminates the collective bias. 

\begin{figure}[tp]
\begin{center}
\includegraphics[width=0.8\hsize]{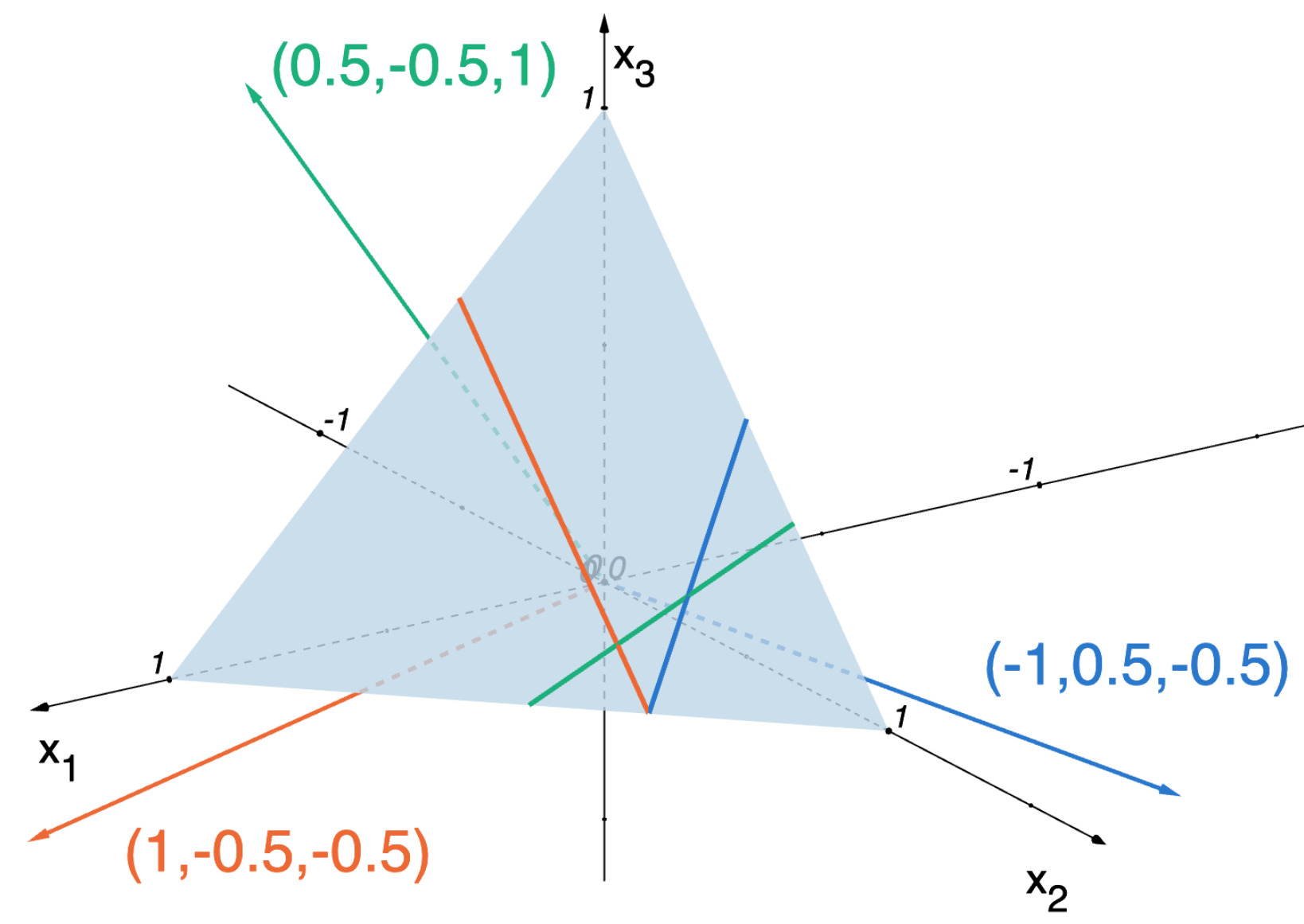}    
\caption{Geometry of the optimal social power sets for different bias vectors. The shaded region is $\Delta_{3}$, on which each colored segment is the set of optimal social power corresponding to the bias vector of the same color.} 
\label{Fig2}
\end{center}
\end{figure}

In particular, Algorithm~\ref{alg:ospca} formulates the pairing problem as an edge cover problem on a complete bipartite graph. Geometrically, $\mathcal{X}$ is a convex polytope whose vertices lie on the edges of $\Delta_n$ and correspond to edges of the complete bipartite graph $\mathcal{G}=(\mathcal{V}^{+},\mathcal{V}^{-},\mathcal{E})$, as illustrated in Fig.~\ref{Fig2}. Specifically, for each undirected edge $\{i,j\}\in\mathcal{E}$, $\frac{\delta_{j}}{\delta_{j}-\delta_{i}}\bd{e}_{i}-\frac{\delta_{i}}{\delta_{j}-\delta_{i}}\bd{e}_{j}$ is a vertex of $\mathcal{X}$. The convex hull of these vertices completely characterizes $\mathcal{X}$, and a minimum edge cover is sufficient to identify an $\bd{x}^{*}\in\interior \Delta_n$. This construction admits a distributed implementation, which will be discussed in Section~\ref{S6}.

\begin{remark}\label{Rde}(Required knowledge of biases)
The optimal social power can be determined without full knowledge of $\bd{\delta}$. Under uniform-sign biases, no knowledge of $\theta$ is required beyond the common sign of $\bd{\delta}$, as discussed after Theorem~\ref{T0}. Under mixed-sign biases, $\mathcal{X}$ is invariant to
replacing $\bd{\delta}$ by $c\bd{\delta}$ with $c\neq 0$. Hence, $\bd{\delta}$ can be measured on benchmark tasks as long as the biases are proportional across tasks in the same domain. This holds, e.g.,
when $\mu_i^{\tau}=(1+s^{\tau}\kappa_i)\theta^{\tau}$ for all tasks $\tau$ in a certain domain, where $\kappa_i$ is agent $i$'s relative bias in this domain and $s^{\tau}\neq 0$ is a task-dependent factor, e.g., reflecting task difficulty. This multiplicative structure is
consistent with the log-normal distribution of estimates observed empirically~\cite{JL-HR-FS-DH:11}. Since $s^{\tau}$ is unknown for a new task, individual estimates cannot
be debiased without per-task calibration, whereas $\mathcal{X}$ and thus the optimal susceptibilities remain the same for all tasks in the same domain. Moreover, the optimal social power is robust to errors in $\bd{\delta}$. For any $\hat{\bd{x}}^{*}$ constructed from measured biases $\hat{\bd{\delta}}$, the collective bias is bounded by the deviation of $\bd{\delta}$ from being proportional to $\hat{\bd{\delta}}$: 
\begin{equation*}
  |\hat{\bd{x}}^{*\top}\bd{\delta}|
  \leq \min_{c\in\mathbb{R}}\max_{i}|\delta_i-c\hat{\delta}_i|,
\end{equation*}
since $\hat{\bd{x}}^{*\top}\bd{\delta}
=\hat{\bd{x}}^{*\top}(\bd{\delta}-c\hat{\bd{\delta}})$
for all $c\in\mathbb{R}$ and $\hat{\bd{x}}^{*\top}\hat{\bd{\delta}}=0$.
\end{remark}

\section{Optimal social power realization}\label{S5}

Solving problem~\eqref{eq11} yields an optimal social power
$\bd{x}^{*}$ over the social power simplex. In this section, we synthesize susceptibility vectors $\bd{a}^{*}$ that realize $\bd{x}^{*}$, thereby solving problem~\eqref{eq3}. The feasibility of realizing $\bd{x}^*$ is characterized by the notions of susceptibility realizability introduced in Definition~\ref{D1}.

By Proposition~\ref{P4}, realizing $\bd{x}^{*}$ amounts to finding a nonnegative delegated power $\bd{\gamma}^{*}$ satisfying $L^{\top}\bd{\gamma}^{*}=\bd{1}_{n}/n-\bd{x}^{*}$, from which~\eqref{eq12a} yields $a^{*}_i=\gamma^{*}_i/(x^{*}_i+\gamma^{*}_i)$. Here, $x^{*}_i+\gamma^{*}_i>0$
since $x^{*}_i+\gamma^{*}_i=1/n+\sum_{j\in\V}W_{ji}\gamma^{*}_j\geq 1/n$ by~\eqref{eq12b}. A natural candidate for $\bd{\gamma}^{*}$ is the minimum-norm least-squares solution
$(L^{\top})^{\dagger}(\bd{1}_{n}/n-\bd{x}^{*})$. However, it is in general neither an exact solution nor nonnegative. The next theorem characterizes FSR and gives all optimal
susceptibilities in closed form.

\begin{theorem}\label{T3}(Susceptibility realizability of social power)
\begin{enumerate}
  \item The FJ model is FSR if and only if $\mathcal{G}(W)$ is strongly connected. \label{T3-1}
  \item Suppose that $\mathcal{G}(W)$ is strongly connected with eigenvector centrality $\bd{\omega}$, and let $\mathcal{X}$ be the set of optimal solutions to problem~\eqref{eq11}. Then, $\bd{a}^{*}$ solves problem~\eqref{eq3} if and only if $\bd{a}^{*}=\dg{\bd{x}^{*}+\bd{\gamma}^{*}}^{-1}\bd{\gamma}^{*}$ with \label{T3-2}
\begin{align}\label{eq:optsus}
  \bd{\gamma}^{*}=(L^{\top})^{\dagger}\bd{b}^{*}+\beta\bd{\omega}
\end{align}
for some $\bd{x}^{*}\in\mathcal{X}$, where $\bd{b}^{*}=\bd{1}_{n}/n-\bd{x}^{*}$ and $\beta\in\mathbb{R}$ is a
  constant ensuring $\bd{\gamma}^{*}\geq 0$.
\end{enumerate}
\end{theorem}

\begin{figure}[tp]
\begin{center}
\includegraphics[width=0.85\hsize]{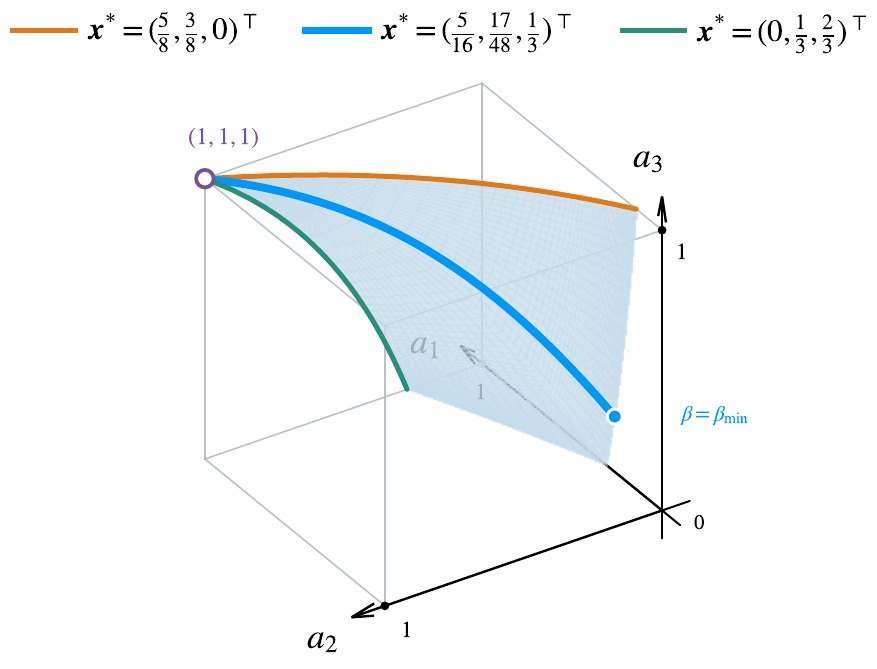}   
\caption{Geometry of the optimal susceptibility set for $W=[0.3,\ 0.15,\ 0.55;\ 0.35,\ 0.35,\ 0.3;\ 0.7,\ 0.05,\ 0.25]$ and $\bd{\delta}=(0.6,\ -1, \ 0.5)^{\top}$. The shaded surface is generated by jointly varying
$\bd{x}^{*}\in\mathcal{X}$ and $\beta$ subject to $\bd{\gamma}^{*}\geq0$ in~\eqref{eq:optsus};
the open marker denotes the excluded limit $\bd{1}_3$ as $\beta\to\infty$. The curves correspond to different $\bd{x}^{*}$, and the marked point on the middle curve is obtained at the minimum feasible
value $\beta=\beta_{\min}$.}\label{Fig2.1}
\end{center}
\end{figure}

Theorem~\ref{T3} is proved in Appendix~\ref{app:T3}. It establishes the equivalence between FSR and strong connectivity. Note that problems~\eqref{eq3} and~\eqref{eq11} have the same minimum if and only if some $\bd{x}^{*}\in\mathcal{X}$ is realizable. Strong connectivity thus guarantees that SPRR solves problem~\eqref{eq3} exactly, yielding all optimal susceptibilities in closed form via the Laplacian pseudoinverse and the eigenvector centrality; otherwise, it may happen that no $\bd{x}^{*}\in\mathcal{X}$ is realizable. 

Combining the closed-form expression~\eqref{eq:optsus} in Theorem~\ref{T3} with the optimal social power set $\mathcal{X}$ in~\eqref{eqosp} and Theorem~\ref{T1} yields a complete characterization of the globally optimal susceptibility set. Fig.~\ref{Fig2.1} illustrates the geometry of this set for a three-agent strongly connected $\mathcal{G}(W)$ and a mixed-sign $\bd{\delta}$. For each $\bd{x}^{*}\in\mathcal{X}$, varying $\beta$ over all values ensuring $\bd{\gamma}^{*}\geq0$ traces a curve of optimal susceptibility vectors that induce the same $\bd{x}^{*}$. As $\bd{x}^{*}$ varies over $\mathcal{X}$, these curves form the entire optimal susceptibility surface for the given $W$ and $\bd{\delta}$. Moreover, along each curve, the smallest feasible $\beta$, denoted by
$\beta_{\min}$, gives the minimum-budget optimal susceptibilities inducing $\bd{x}^{*}$. At the other extreme, letting $\beta\to\infty$ in~\eqref{eq:optsus} yields $\bd{\gamma}^{*}\to\infty$ and hence
$\bd{a}^{*}\to\bd{1}_n$. Thus, achieving collective wisdom does not necessarily require a pronounced separation between stubborn holdouts and susceptible swingers: the minimum collective bias remains attainable even in a group whose members are all highly receptive to others' opinions.

\begin{remark}(Interpretation of optimal susceptibilities) 
In view of Remark~\ref{R:redis}, $(L^{\top})^{\dagger}\bd{b}^{*}$ in~\eqref{eq:optsus} is the minimum-norm delegation that redistributes power from $\bd{1}_{n}/n$ to $\bd{x}^{*}$. Since $L^{\top}\bd{\omega}=0$, $\beta\bd{\omega}$ is a circulation in which each agent receives exactly what it delegates: it leaves social power unchanged and only increases delegation, which
ensures $\bd{\gamma}^{*}\geq 0$. A larger $\beta$ makes each agent delegate a larger fraction $a_i^{*}$ of its incoming power, which explains $\bd{a}^{*}\to\bd{1}_n$ as $\beta\to\infty$.
\end{remark}

Although strong connectivity is necessary for FSR, it is not
necessary for $\bd{x}$-SR. The following corollary, which follows from the proof of Theorem~\ref{T3}, characterizes the necessary and sufficient conditions for $\bd{x}$-SR.

\begin{corollary}\label{C0}
Let $\bd{x}\in\Delta_{n}$ and $\bd{b}=\bd{1}_{n}/n-\bd{x}$. Then,
\begin{enumerate}
  \item the FJ model is $\bd{{x}}$-SR if and only if \label{C0-1}
  \begin{enumerate}
  \item $\bd{b}\in \mathbf{Null}(L)^{\perp}$, and \label{cond1}
  \item $\zeta_{i}\geq 0$ for all $i\in\mathcal{T}$, where 
  $\bd{\zeta }=(L^{\top})^{\dagger}\bd{b}$ and $\mathcal{T}$
  is the set of agents not in any sink SCC of $\mathcal{G}(W)$;\label{cond2}
  \end{enumerate}
  
  \item \label{C0-2} suppose that $\bd{x}^{*}\in\mathcal{X}$ and the FJ model is $\bd{{x}}^{*}$-SR. Then, $\bd{a}^{*}$ induces $\bd{x}^{*}$, and hence solves problem~\eqref{eq3}, if and only if $\bd{a}^{*}=\dg{\bd{x}^{*}+\bd{\gamma}^{*}}^{-1}\bd{\gamma}^{*}$ with
  \begin{align}\label{eq:optsus2}
    \bd{\gamma}^{*}=(L^{\top})^{\dagger}\bd{b}^{*}+\bd{\omega}^{\prime},
  \end{align}
  where $\bd{b}^{*}=\bd{1}_{n}/n-\bd{x}^{*}$ and $\bd{\omega}^{\prime}$ is a left eigenvector of $W$ associated with $1$ that ensures $\bd{\gamma}^{*}\geq 0$. 
\end{enumerate}
\end{corollary}

In Corollary~\ref{C0}, condition~\ref{cond1} ensures the existence of a solution to $L^{\top}\bd{\gamma}=\bd{b}$, while condition~\ref{cond2} guarantees its nonnegativity. In particular,~\ref{cond1} 
explains how the connectivity of $\mathcal{G}(W)$ constrains susceptibility realizability.
Assume that $\mathcal{G}(W)$ has $m$ sink SCCs, so that $L$ has the block form 
\begin{align*}
L=\begin{bmatrix}
   L_{1}  & \cdots & 0& 0\\
    \vdots   & \ddots & \vdots & \vdots\\
    0 &  \cdots & L_{m}& 0\\
    L_{m+1,1} & \cdots & L_{m+1,m}& L_{m+1}
\end{bmatrix},
\end{align*}
where $L_{m+1}$ is associated with $\mathcal{T}$ and $L_{i}\in\mathbb{R}^{n_{i}\times n_{i}}$. Then, for each $i\leq m$, there exists $\bd{d}^{i}\in\mathbb{R}^{n_{m+1}}$ such that $L_{m+1,i}\bd{1}_{n_i}+L_{m+1}\bd{d}^{i}=0$ and $\sum_{i=1}^{m}\bd{d}^{i}=\bd{1}_{n_{m+1}}$. Therefore, $\mathbf{Null}(L)$ is spanned by $\bd{v}^{1},\dots,\bd{v}^{m}$, where  
\begin{align*}
\bd{v}^{i}
=[\bd{0}_{n_1}^{\top},\ldots,\bd{0}_{n_{i-1}}^{\top}, \ \bd{1}_{n_i}^{\top}, \ \bd{0}_{n_{i+1}}^{\top},\ldots,\bd{0}_{n_m}^{\top}, \
(\bd{d}^{i})^{\top}]^{\top}.
\end{align*}
As a result,~\ref{cond1} requires $(\bd{1}_{n}/n-\bd{x})^{\top}\bd{v}^{i}=0$ for all $i\leq m$.
This always holds for $m=1$, but becomes increasingly restrictive as $m$ increases. Thus, stronger connectivity tends to yield stronger realizability, as illustrated in Example~\ref{exa2}.

\begin{example}(Realizability and connectivity)\label{exa2}
Fig.~\ref{fig3} illustrates the susceptibility realizability of the FJ model. Fig.~\ref{fig3}(a) shows uniformly sampled points of $\bd{a}\in[0,1)^{3}$, colored by their RGB coordinates. For each $\bd{a}$, we 
compute the induced social power $\bd{x}$ and plot it in $\Delta_{3}$ using the same color, under influence matrices $W_1$, $W_2$, and $W_3$ given by
\begin{align*}
\begin{bmatrix}
  0.5 & 0.5 & 0\\
  0 & 0.5 & 0.5\\
  0.5 & 0 & 0.5
\end{bmatrix}, \quad \begin{bmatrix}
  0.5 & 0.5 & 0\\
  0.5 & 0.5 & 0\\
  0.5 & 0 & 0.5
\end{bmatrix}, \quad \begin{bmatrix}
  0.5 & 0.5 & 0\\
  0.5 & 0.5 & 0\\
  0 & 0 & 1
\end{bmatrix},
\end{align*}
respectively. Here, $\mathcal{G}(W_1)$ is strongly connected, $\mathcal{G}(W_2)$ 
has two globally reachable nodes, and 
$\mathcal{G}(W_3)$ contains an isolated node. As shown in Figs.~\ref{fig3}(b)--(d), the induced social power covers the interior of $\Delta_{3}$ under $W_1$, a proper subset under $W_2$, and a lower-dimensional subset under $W_3$, consistent with Theorem~\ref{T3}~\ref{T3-1}, Corollary~\ref{C0}, and the preceding discussion.
  \begin{figure}
\centering

\begin{minipage}{0.47\linewidth}
    \centering
    \includegraphics[width=\linewidth]{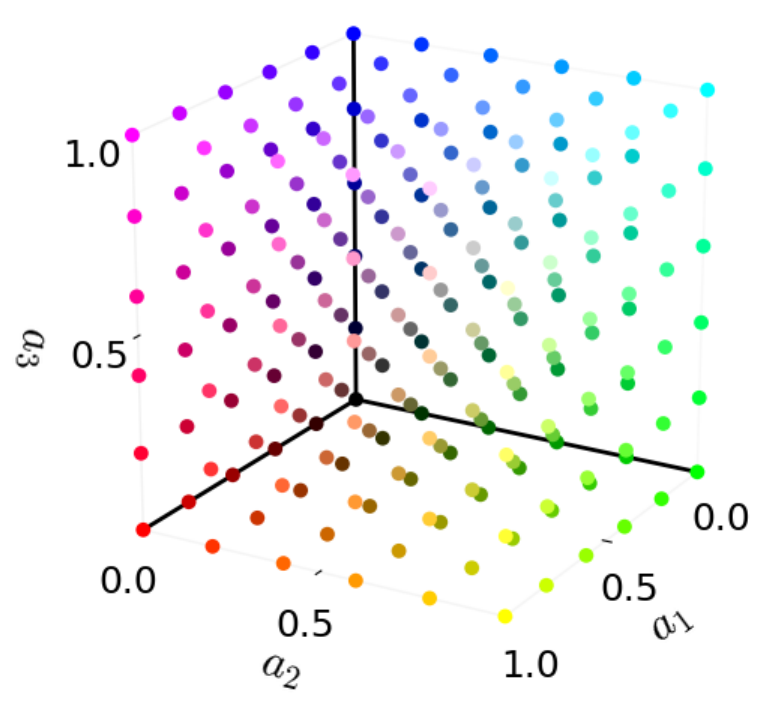}
    
    (a) 
\end{minipage}
\hfill
\begin{minipage}{0.47\linewidth}
    \centering
    \includegraphics[width=\linewidth]{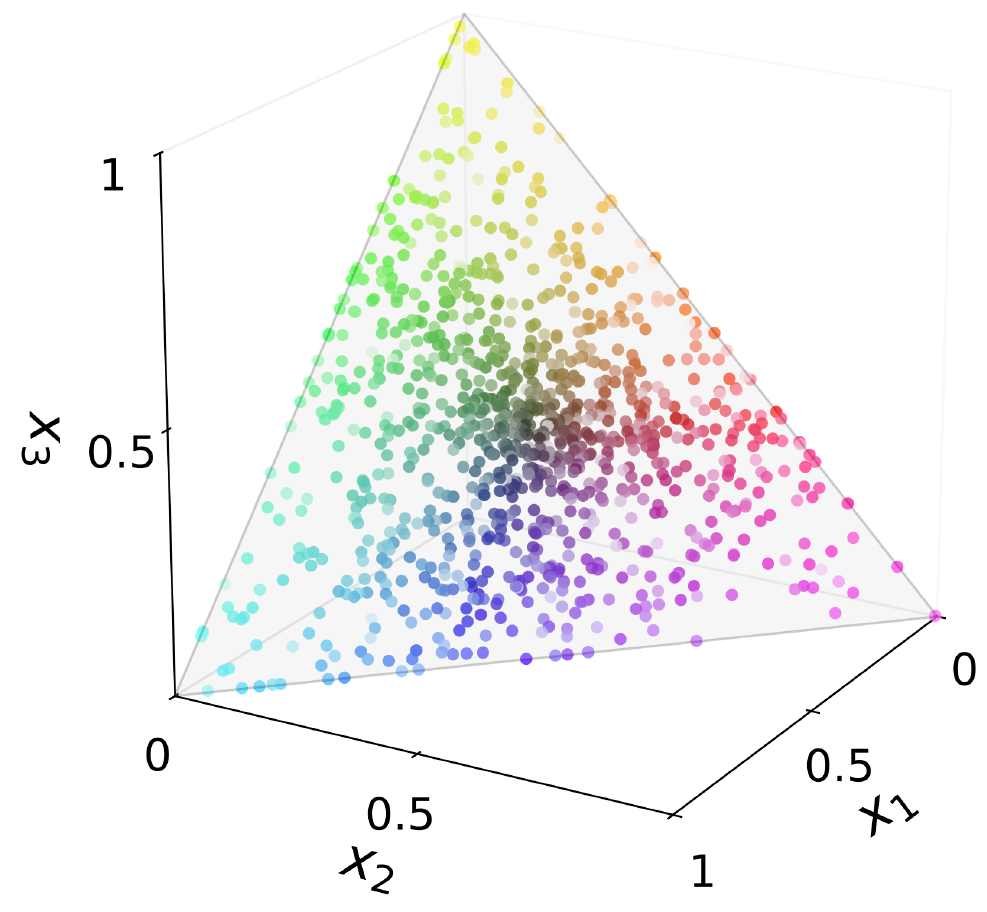}
    
    (b)
\end{minipage}

\vspace{0.1cm}

\begin{minipage}{0.47\linewidth}
    \centering
    \includegraphics[width=\linewidth]{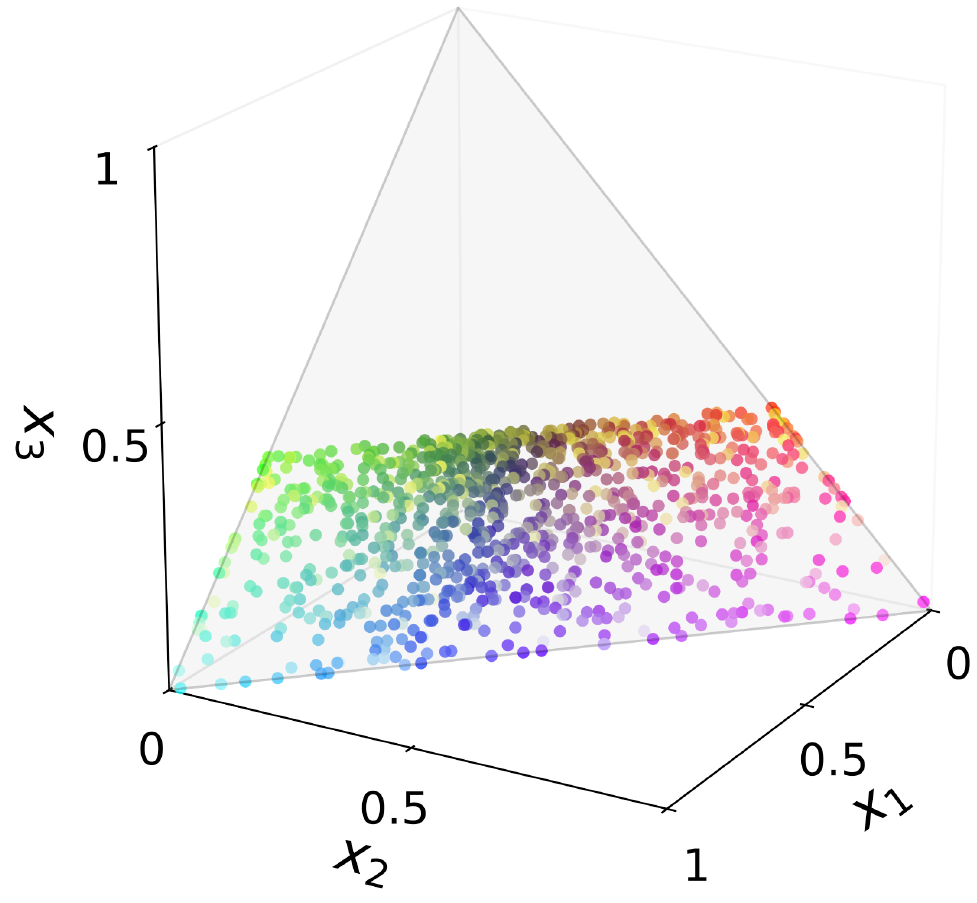}
    
    (c) 
\end{minipage}
\hfill
\begin{minipage}{0.47\linewidth}
    \centering
    \includegraphics[width=\linewidth]{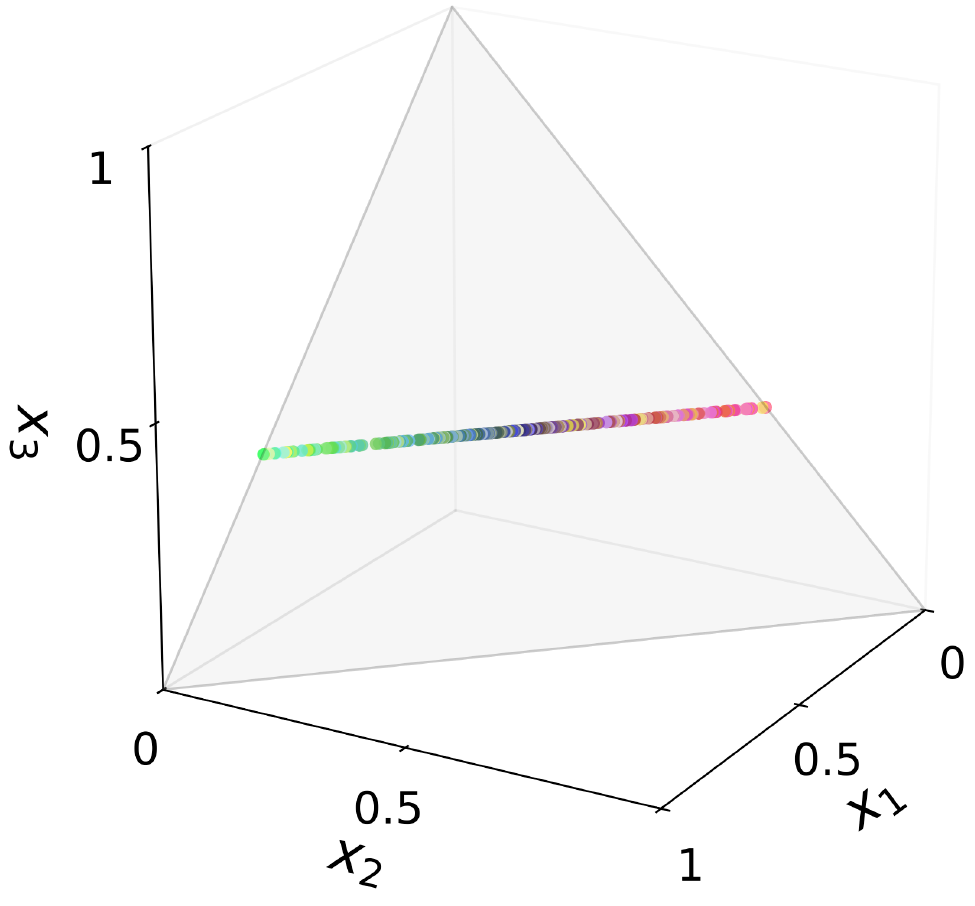}
    
    (d) 
\end{minipage}

\caption{Example~\ref{exa2}. Illustration of susceptibility realizability of the FJ model with $n=3$. (a) Points of $\bd{a}$ uniformly sampled from $[0,1)^{3}$ colored by RGB value, i.e., $(a_1,a_2,a_3)\to (R,G,B)$. (b)--(d) Social power $\bd{x}$ in $\Delta_{3}$, colored by the RGB value of the associated $\bd{a}$, for a strongly connected network (b), a network with two globally reachable nodes (c), and a network
containing an isolated node (d).}
\label{fig3}
\end{figure}
\end{example} 

Example~\ref{exa2} shows that weaker connectivity shrinks the set of realizable social power. In particular, the FJ model may fail to be $\bd{x}^{*}$-SR for every $\bd{x}^{*}\in\mathcal{X}$, e.g., under uniform-sign biases when some sink SCC contains no agent in $\mathcal{I}_{\min}^{\bd{\delta}}$. This case can still be handled within SPRR by restricting the reformulation to realizable social power characterized in Corollary~\ref{C0}~\ref{C0-1}:
\begin{equation}\label{eq:QPr}
  \begin{aligned}
    \min_{\bd{x}\in\Delta_n}\ \ & g(\bd{x})\\
    \textup{s.t.}\ \ \ & \bd{b}\in\mathbf{Null}(L)^{\perp} \ \textup{and} \ \zeta_i\geq 0,\ \forall i\in\mathcal{T}.
  \end{aligned}
\end{equation}
Since $\bd{b}$ and $\bd{\zeta}$ are affine in $\bd{x}$,~\eqref{eq:QPr} is a convex QP, which is feasible at $\bd{x}=\bd{1}_n/n$ and thus always admits a solution $\bd{x}^{\circ}$. By Corollary~\ref{C0}~\ref{C0-2}, $\bd{x}^{\circ}$ is induced by $\bd{a}^{\circ}$ obtained from~\eqref{eq:optsus2}. Hence, SPRR solves problem~\eqref{eq3} globally for any $\mathcal{G}(W)$. Compared with the realizable case, only $\bd{x}^{\circ}$
requires solving a convex QP, while $\bd{a}^{\circ}$ still follows in closed form. 

Note that~\eqref{eq:QPr} is the convex QP after Proposition~\ref{P4} with $\bd{\gamma}$ eliminated by Corollary~\ref{C0}~\ref{C0-1}. Since the map from $(\bd{x},\bd{\gamma})$ to $\bd{a}$ given
by~\eqref{eq12a} is a continuous bijection, every local minimizer of problem~\eqref{eq3} is global. Moreover, any additional constraint on $\bd{a}$ can be expressed in terms of $(\bd{x},\bd{\gamma})$. In particular, bounds $\bd{l}\leq\bd{a}\leq\bd{u}$ become the linear constraints
$\dg{\bd{l}}(\bd{x}+\bd{\gamma})\leq\bd{\gamma}\leq\dg{\bd{u}}(\bd{x}+\bd{\gamma})$, so that problem~\eqref{eq3} under bounded susceptibilities can also be
solved by a convex QP.
Based on the above results, we summarize SPRR in Algorithm~\ref{alg:sspa}.

\begin{algorithm}[ht]
\caption{SPRR Algorithm}
\label{alg:sspa}
\begin{algorithmic}[1]
  \REQUIRE Influence matrix $W$ and bias vector $\bd{\delta}$ 
\ENSURE Optimal susceptibility $\bd{a}^{*}$

    \STATE Pick $\bd{x}^{*} \in \mathcal{X}$
   \algorithmiccomment{\eqref{eqosp} or Algorithm~\ref{alg:ospca}}
    \STATE Compute $L^{\dagger}$ and identify the sink SCCs of $\mathcal{G}(W)$
    \algorithmiccomment{Tarjan's Algorithm~\cite{RT:72}}
    \IF{the FJ model is not $\bd{x}^{*}$-SR}
      \STATE{ $\bd{x}^{*}\gets$ a solution of~\eqref{eq:QPr}}
    \ENDIF
    \STATE Compute $\bd{\omega}^{\prime}$ with
    $L^{\top}\bd{\omega}^{\prime}=0$ and
    $\omega^{\prime}_i>0$ for all $i\notin\mathcal{T}$
    \STATE $\bd{\gamma}^{*} \!\gets\! 
        (L^{\top})^{\dagger}\big(\frac{\bd{1}_{n}}{n}-\bd{x}^{*}\big)\!+\!\beta \bd{\omega}^{\prime}$
        \algorithmiccomment{pick $\beta$ such that $\bd{\gamma}^{*} \ge 0$}
    \STATE $\bd{a}^{*}\gets\dg{\bd{x}^{*}+\bd{\gamma}^{*}}^{-1}\bd{\gamma}^{*}$
 \RETURN $\bd{a}^{*}$
\end{algorithmic}
\end{algorithm}

\begin{proposition}\label{P41}
Suppose that the FJ model is $\bd{x}^*$-SR for the $\bd{x}^*$ picked in Step $1$ of Algorithm~\ref{alg:sspa}. Then, Algorithm~\ref{alg:sspa} solves problem~\eqref{eq3} with worst-case complexity $\mathcal{O}(n^3)$.
\end{proposition}
\begin{proof}
Correctness follows from Theorem~\ref{T3} and Corollary~\ref{C0}. Step~$4$ is skipped. In Step $2$, computing 
$L^{\dagger}$ requires $\mathcal{O}(n^3)$ time using direct 
methods such as singular value decomposition (SVD), and Tarjan's algorithm~\cite{RT:72} runs in $\mathcal{O}(n+|\mathcal{E}|)$ time, where $|\mathcal{E}|=n^2$ in the dense case. Steps $3$ and $5$ require computing the right and left eigenvector spaces of $W$ associated with $1$, which can be carried out in $\mathcal{O}(n^3)$ time in the worst case. The remaining steps require at most $\mathcal{O}(n^2)$ time. Hence, the overall worst-case complexity is $\mathcal{O}(n^3)$.
\end{proof}

If the FJ model is not $\bd{x}^{*}$-SR for the $\bd{x}^{*}$ picked in Step~$1$, Step~$4$ solves the convex QP~\eqref{eq:QPr} in polynomial time, which returns a realizable $\bd{x}^{*}\in\mathcal{X}$ whenever one exists, and $\bd{x}^{\circ}$ otherwise. Alternatively, under mixed-sign biases, another $\bd{x}^{*}\in\mathcal{X}$ can be generated
without solving~\eqref{eq:QPr} by modifying the edge cover
and the convex weights $\bd{z}$ in Algorithm~\ref{alg:ospca}, as illustrated in the following example.

\begin{example}(Performance of SPRR Algorithm)\label{exa3}
Consider a strongly connected $\mathcal{G}(W)$ consisting of $5$ agents with 
\begin{align*}
W=\begin{bmatrix}
  0.1 & 0.6 & 0.3 & 0 & 0\\
  0 & 0.2& 0.5 & 0.3 & 0\\
  0 & 0 & 0.3 & 0.4 & 0.3\\
  0.2 & 0 & 0 & 0.2 & 0.6\\
  0.5 & 0.2 & 0 & 0 & 0.3
  \end{bmatrix}. 
\end{align*}
Let $\theta=0$ and $\bd{\delta}=\bd{\mu}=(-2, 5, 3,-6,1)^{\top}$. 
We first construct an optimal $\bd{x}^{*}$ using Algorithm~\ref{alg:ospca}. Let $\mathcal{V}^{+}=\{2,3,5\}$ and $\mathcal{V}^{-}=\{1,4\}$. A minimum edge cover over the complete bipartite graph $\mathcal{G}=(\mathcal{V}^{+},\mathcal{V}^{-},\mathcal{E})$ is $\{1,2\},\{3,4\}, \{1,5\}$. With $q(1)=2$, taking the uniform convex weights $\bd{1}_{5}/5$ yields $\bd{x}^{*}=(\frac{37}{105},\ \frac{4}{35},\ \frac{4}{15},\ \frac{2}{15},\ \frac{2}{15})^{\top}$.
By Theorem~\ref{T3}, setting $\beta$ to its minimum feasible value $\beta_{\min}\approx 0.61$
gives $\bd{a}^{*}\approx(0,\ 0.59,\ 0.08,\ 0.54,\ 0.64)^{\top}$.
Moreover, any $\beta> \beta_{\min}$ yields another optimal susceptibility vector inducing the same $\bd{x}^{*}$. 
Fig.~\ref{Fig4}(a) shows the trajectories of $y_{\textup{ave}}(k)$ under $\bd{y}(0)=\bd{\mu}$ with various choices of $\beta$. The baseline susceptibility $\bd{a}=\bd{1}_{5}/2$ is not optimal, leading to a biased trajectory. Consistent with our theoretical results, all optimal susceptibilities drive $y_{\textup{ave}}(k)$ to $0$, with slower convergence as $\beta$ increases.  
\begin{figure}
\centering

\begin{minipage}{0.49\linewidth}
    \centering
    \includegraphics[width=\linewidth]{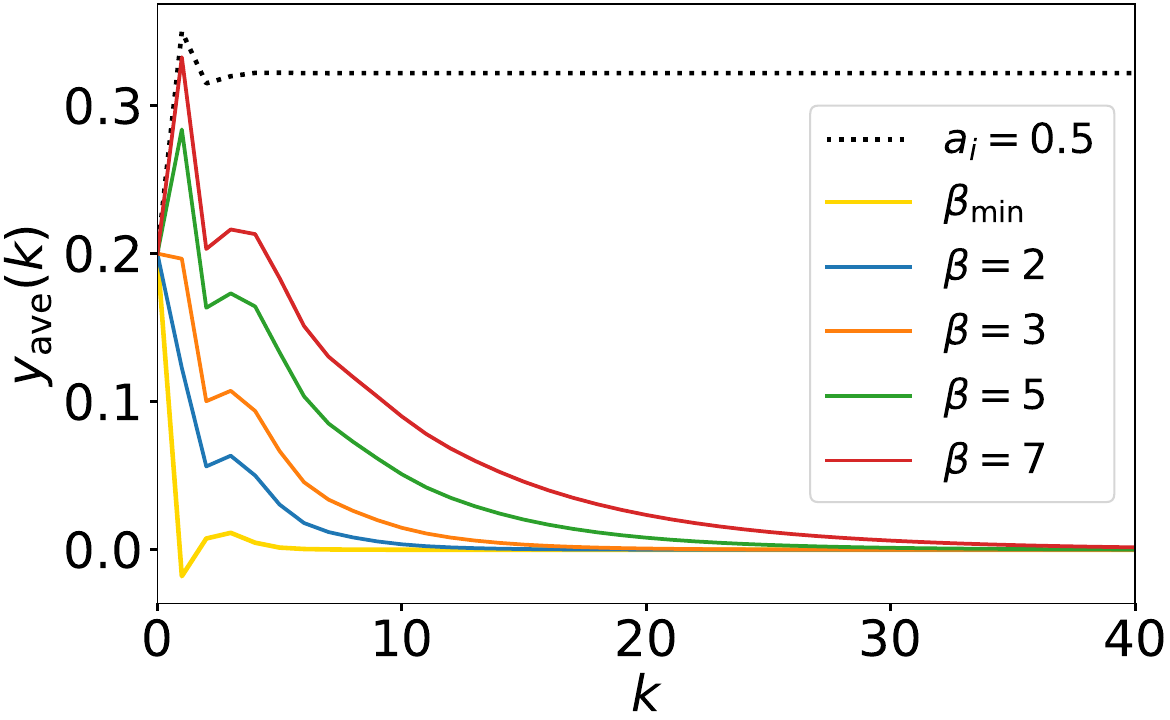}
    
    (a) $\mathcal{G}(W)$
\end{minipage}
\hfill
\begin{minipage}{0.495\linewidth}
    \centering
    \includegraphics[width=\linewidth]{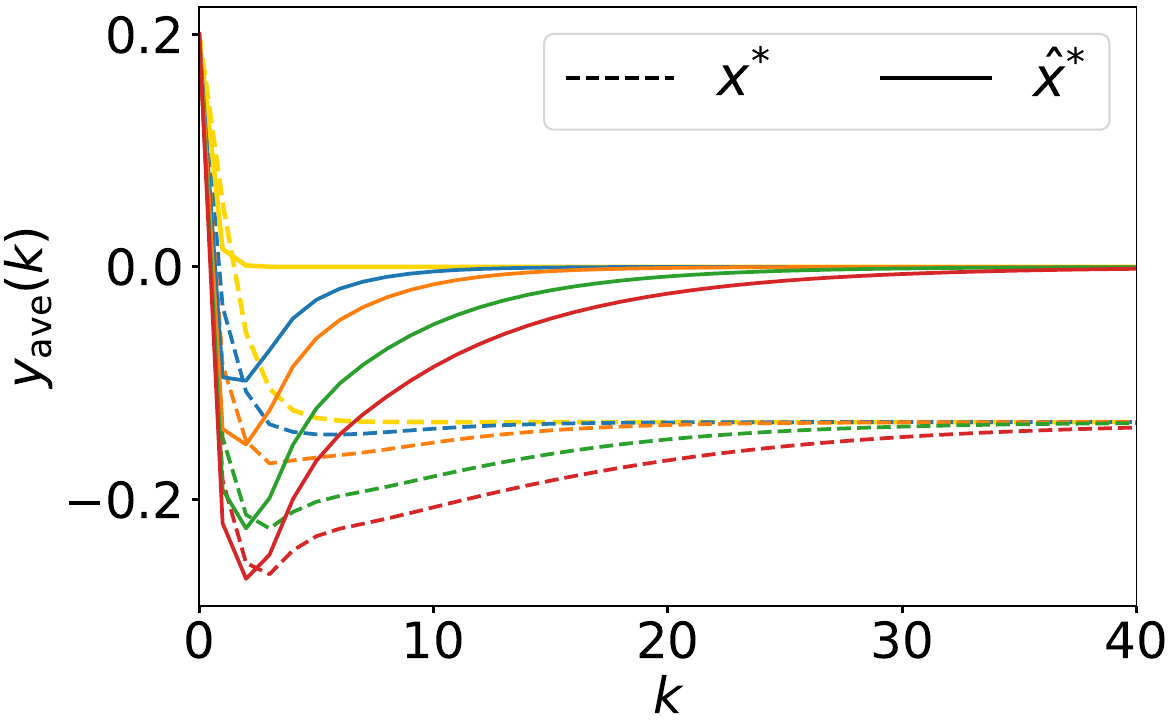}
    
    (b) $\mathcal{G}(\hat{W})$
\end{minipage}
\caption{Example~\ref{exa3}. Trajectories of $y_{\textup{ave}}(k)$ under different influence networks and susceptibilities.} 
\label{Fig4}
\end{figure}

Next, consider $\mathcal{G}(\hat{W})$ consisting of $2$ sink SCCs with 
\begin{align*}
\hat{W}=\begin{bmatrix}
  0.7 & 0.3 & 0 & 0 & 0\\
  0 & 0.5& 0.5 & 0 & 0\\
  0.6 & 0 & 0.4 & 0 & 0\\
  0 & 0 & 0 & 0.4 & 0.6\\
  0 & 0 & 0 & 0.7 & 0.3
  \end{bmatrix},
\end{align*}
under which the FJ model is not $\bd{x}^{*}$-SR for the above $\bd{x}^{*}$ due to violation of condition~\ref{cond1} in Corollary~\ref{C0}. 
The dashed lines in Fig.~\ref{Fig4}(b) show the trajectories of $y_{\textup{ave}}(k)$ with $\bd{a}$ computed from $\bd{x}^{*}$ using~\eqref{eq:optsus2},
where $y_{\textup{ave}}(k)$ does not converge to $0$. 
To restore realizability, we compute an alternative optimal social power $\tilde{\bd{x}}^{*}=(\frac{21}{95},\ \frac{18}{95},\ \frac{18}{95},\ \frac{20}{95},\ \frac{18}{95})^{\top}$ using a different edge cover and convex weights in Algorithm~\ref{alg:ospca}, for which the FJ model is $\tilde{\bd{x}}^{*}$-SR. We compute $\bd{a}^{*}$ with $\bd{\omega}^{\prime}=\beta (\hat{W}^{\infty})^{\top}\bd{1}_{n}/n$. As shown by the solid lines
in Fig.~\ref{Fig4}(b), the trajectories of $y_{\textup{ave}}(k)$ with $\bd{a}^{*}$ converge to $0$ under different choices of $\beta$.  
\end{example}

\begin{remark}(SPRR for other collective tasks)\label{R:variance}
By~\eqref{eqmv}, problem~\eqref{eqMSE} can be rewritten over the social power simplex as 
\begin{equation}\label{mvMSE}
\min_{\bd{x}\in\mathcal{X}}\ \sum_{i=1}^{n}\sigma_i^{2}x_i^{2},
\end{equation}
which is a strictly convex QP, and hence admits a unique solution $\bd{x}^{*}$. 
If the FJ model is not $\bd{x}^{*}$-SR, $\mathcal{X}$ in~\eqref{mvMSE} is replaced by the constraints in~\eqref{eq:QPr} together with $\bd{\delta}^{\top}\bd{x}=\bd{\delta}^{\top}\bd{x}^{\circ}$, where $\bd{x}^{\circ}$ is a solution of~\eqref{eq:QPr}, so that~\eqref{mvMSE} remains a strictly convex QP. In either case, the optimal susceptibilities follow from Corollary~\ref{C0}. The MSE in~\eqref{MSE} can be minimized within SPRR similarly. More broadly, whenever a collective task prescribes a target social power, Corollary~\ref{C0} determines whether it is realizable and, if so,
yields the susceptibilities inducing it, for instance, the social power allocations shaping collective wisdom under unbiased estimates~\cite{FB-FF-BF:20, YT-LW-FB:23}.
\end{remark}

In practice, instead of using the computationally expensive SVD, $L^{\dagger}$ can be computed using graph sparsification techniques and iterative methods~\cite{Saad2003}. Nonetheless, most of them require centralized knowledge of $L$. This motivates the distributed iterative 
algorithms proposed in the next section, which enable fully decentralized computation of $\bd{a}^{*}$.

\section{Decentralized implementation of SPRR}\label{S6}

While SPRR characterizes the complete set of optimal
solutions to problem~\eqref{eq3}, Algorithm~\ref{alg:sspa} relies on global information, e.g., to construct the edge cover and compute $L^{\dagger}$. Nevertheless, Theorem~\ref{T3} implies that computing $\bd{a}^{*}$ does not require $L^{\dagger}$; it suffices to compute a particular solution of $L^{\top}\bd{z}=\bd{b}^{*}$. 
In this section, we develop a distributed implementation of SPRR in which each agent determines its own optimal susceptibility through only local interactions.

Let $\mathcal{N}_{i}^{+}=\setdef{j}{W_{ji}>0}$ and
$\mathcal{N}_{i}^{-}=\setdef{j}{W_{ij}>0}$ denote the sets of $i$'s in- and out-neighbors in $\mathcal{G}(W)$. For simplicity, we focus on strongly connected $\mathcal{G}(W)$. Suppose that each agent has access only to its own bias and the biases of its immediate neighbors. Under uniform-sign biases, the decentralized implementation is immediate. By~\eqref{opti0}, $\bd{a}^{*}$ solves problem~\eqref{eq3} if and only if $a^{*}_{i}=1$ for all $i\notin\mathcal{I}_{\min}^{\bd{\delta}}$ and $a^{*}_{j}<1$ for some $j\in\mathcal{I}_{\min}^{\bd{\delta}}$. Since
$\mathcal{I}_{\min}^{\bd{\delta}}=\argmin_{j}\mu_j$, each agent can determine whether it belongs to $\mathcal{I}_{\min}^{\bd{\delta}}$ via the min-consensus
$\vartheta_{i}(k+1)=\min_{j\in\mathcal{N}_{i}^{-}\cup\{i\}}\vartheta_{j}(k)$ with $\vartheta_i(0)=\mu_i$, which reaches $\min_{j}\mu_j$ within $n-1$ steps. Then, agent $i$ sets $a^{*}_{i}=1$ if $\mu_i>\vartheta_i(n-1)$, and any $a^{*}_{i}\in[0,1)$ otherwise.

Under mixed-sign biases, the main obstacle to decentralizing Algorithm~\ref{alg:ospca} is the construction of the edge cover.
Given the one-hop local information structure, Algorithm~\ref{alg:ospca} can be decentralized if and only if the following assumption holds. 

\begin{assumption}\label{A4}
For all $i\in\V$, there exists $j\in\mathcal{N}_{i}^{-}\cup\mathcal{N}_{i}^{+}$ such that $\delta_{i}\delta_{j}<0$.
\end{assumption}

Under Assumption~\ref{A4}, each agent $i$ can select a neighbor $j$ with $\delta_{i}\delta_{j}<0$ to compute the mixing weight $\eta_i$ as in Step~$3$ of Algorithm~\ref{alg:ospca}, and shares $\eta_i$ with $j$; $j$ may either pair with $i$ or with another neighbor. This local pairing structure induces an edge cover of the complete bipartite graph $\mathcal{G}=(\mathcal{V}^{+},\mathcal{V}^{-},\mathcal{E})$ defined in Step~$2$ of Algorithm~\ref{alg:ospca}. We next consider distributed susceptibility realization. 

\begin{theorem}\label{T4}(Distributed susceptibility realization of social power)
Suppose that $\mathcal{G}(W)$ is strongly connected with eigenvector centrality $\bd{\omega}$ and $\bd{x}^{*}\in\mathcal{X}$. Let $\bd{b}^{*}=\bd{1}_{n}/n-\bd{x}^{*}$ and $M_{\alpha}=\alpha W^\top+(1-\alpha)I_{n}$ with $\alpha\in (0,1)$. Then, 
\begin{enumerate}
  \item  \label{T4-1}system 
  \begin{align}\label{eqomega}
    \tilde{\bd{\omega}}(k+1)=M_{\alpha}\tilde{\bd{\omega}}(k)
  \end{align}
  exponentially converges to $\bd{\omega}$ if and only if $\tilde{\bd{\omega}}^{\top}(0)\bd{1}_{n}=1$;
  
\item \label{T4-2}for any $\bd{z}(0)\in\mathbb{R}^{n}$, system 
  \begin{align}\label{eqz}
    \bd{z}(k+1)=M_{\alpha}\bd{z}(k)+\alpha \bd{b}^{*}
  \end{align}
  exponentially converges to a solution $\bd{z}^{*}$ of $L^{\top}\bd{z}=\bd{b}^{*}$; 

\item \label{T4-3} $\bd{a}^{*}=\dg{\bd{x}^{*}+\bd{z}^{*} + \beta\bd{\omega}}^{-1} (\bd{z}^{*} +  \beta\bd{\omega})$ solves problem~\eqref{eq3} for all $\beta\geq \max_{i}-z^{*}_{i}/\omega_{i}$;
\item \label{T4-4} $\bd{a}^{*}=\dg{\bd{x}^{*}+\bd{z}^{*}}^{-1}\bd{z}^{*}$ solves problem~\eqref{eq3} if
  \begin{align}\label{eqz01}
    \bd{z}^{\top}(0)\bd{1}_{n}\geq \max_{i}-\alpha\frac{\phi_{i} }{\omega_{i}},
  \end{align}
  where $\bd{\phi}=(I_{n}-U)^{-1}\bd{b}^{*}$ with $U=M_{\alpha}-\bd{\omega}\bd{1}_{n}^{\top}$.
\end{enumerate}
\end{theorem}

Theorem~\ref{T4}, proved in Appendix~\ref{app:T4}, shows that $\bd{a}^{*}$ can be computed using fully distributed linear iterations, in which each agent $i$ updates $\tilde{\omega}_{i}(k)$ and $z_{i}(k)$ using only information from its in-neighbors $j\in\mathcal{N}_{i}^{+}$. Note that statements~\ref{T4-1} and~\ref{T4-2} hold as long as $\mathcal{G}(W)$ has a unique sink SCC. Strong connectivity is imposed only to ensure $\bd{\omega}>0$ in~\ref{T4-3} and~\ref{T4-4}; these results can be extended to the case of a unique sink SCC, provided that condition~\ref{cond2} of Corollary~\ref{C0} holds for $\bd{x}^{*}$, or with $\bd{x}^{*}$ replaced by $\bd{x}^{\circ}$ in~\eqref{eq:QPr} when no $\bd{x}^{*}\in\mathcal{X}$ is realizable. Finally,~\ref{T4-4} removes the need for explicit use of $\bd{\omega}$ at the cost of a more restrictive initialization.
Since the optimal susceptibilities under uniform-sign biases can be obtained directly via the min-consensus above, we summarize the decentralized SPRR for mixed-sign biases in Algorithm~\ref{alg:dsspa}. Since multiplying $\bd{x}^*$, $\bd{b}^*$, $\bd{z}(0)$, and $\tilde{\bd{\omega}}(0)$ by $n$ leaves $\bd{a}^*$ unchanged, Algorithm~\ref{alg:dsspa} can be implemented without knowledge of $n$.

\begin{algorithm}
\caption{Decentralized SPRR (dSPRR) Algorithm}
\label{alg:dsspa}
\begin{algorithmic}[1]
  \REQUIRE Influence matrix $W$, bias vector $\bd{\delta}$, and stopping time $T>0$
\ENSURE Approximate optimal susceptibility $\hat{\bd{a}}^{*}$
\STATE Set $\alpha\in (0,1)$, and initialize $\bd{z}(0) \in \mathbb{R}^n$ and $\tilde{\bd{\omega}}(0)=\bd{1}_{n}/n$

\STATE For each $i\in\V$, pick a $q(i)\in\mathcal{N}_{i}^{-}\cup\mathcal{N}_{i}^{+}$ with $\delta_i\delta_{q(i)}<0$, compute $\eta_i \gets \frac{\delta_{q(i)}}{\delta_{q(i)} - \delta_i}$ and send $\eta_i$ to $q(i)$

\STATE For each $i\in\V$, compute 
\begin{align*}
  x_i^* \gets \frac{1}{n} (\eta_i + \sum_{j\in\{l|q(l)=i\}}(1-\eta_j)) \quad \textup{and} \quad b_i^*\gets \frac{1}{n}-x_i^*
\end{align*}

\FOR{$k = 0$ to $T-1$}  
\STATE For each $i\in\V$,
\begin{align*}
  z_{i}(k+1)&=\alpha\sum_{j\in\mathcal{N}_{i}^{+}}W_{ji}z_{j}(k)+(1-\alpha)z_{i}(k)+\alpha b_i^*, \\
  \tilde{\omega}_{i}(k+1)&=\alpha\sum_{j\in\mathcal{N}_{i}^{+}}W_{ji}\tilde{\omega}_{j}(k)+(1-\alpha)\tilde{\omega}_{i}(k)
\end{align*}
\ENDFOR
\STATE For each $i\in\V$, $\vartheta_i(0)\gets -z_i(T)/\tilde{\omega}_i(T)$
\FOR{$k = 0$ to $T-1$}  
\STATE For each $i\in\V$, $\vartheta_{i}(k+1)=\max_{j\in\mathcal{N}_{i}^{-}\cup\{i\}}\vartheta_{j}(k)$ 
\ENDFOR
\STATE For each $i\in\V$, $\chi_i\gets z_i(T)+\vartheta_i(T)\tilde{\omega}_i(T)$ and $\hat{a}^{*}_i\gets\chi_i/(x^{*}_i+\chi_i)$
\RETURN $\hat{\bd{a}}^{*}$
\end{algorithmic}
\end{algorithm}

\begin{proposition}\label{P5}
Suppose that Assumption~\ref{A4} holds, and $\mathcal{G}(W)$ is strongly connected with $|\mathcal{E}|$ edges. For Algorithm~\ref{alg:dsspa} with output $\hat{\bd{a}}^{*}$ after $T$ iterations,
\begin{enumerate}
\item  \label{P5-1}the total communication complexity is $\mathcal{O}(T|\mathcal{E}|)$;

\item \label{P5-2} let $\bd{a}^{*}$ be as in Theorem~\ref{T4}~\ref{T4-3} with 
$\beta^{*}=\max_{i}(-z_{i}^{*}/\omega_{i})$, where $\bd{z}^{*}$ is the limit of~\eqref{eqz} from the same $\bd{z}(0)$. Then, for any $\epsilon>0$, there exists $T_{\epsilon}>0$ such that $\norm{\hat{\bd{a}}^{*}-\bd{a}^{*}}_{2}\leq\epsilon$ for all $T>T_{\epsilon}$.
\end{enumerate}
\end{proposition}

Proposition~\ref{P5} is proved in Appendix~\ref{app:T4}. Theorem~\ref{T4} and Proposition~\ref{P5} show that SPRR can be implemented in a fully distributed manner. In particular, the communication cost of the dSPRR Algorithm scales linearly with the number of edges, making it suitable for large-scale networked systems. Moreover, this decentralized implementation provides a mechanistic account of how collective intelligence can emerge spontaneously in practice. 
As long as each agent is aware of the expertise of at least one neighbor with an opposing viewpoint (i.e., an opposite bias in the task domain), it can determine its own optimal susceptibility through repeated local interactions with its neighbors. Collectively, these local adaptations yield an optimal susceptibility vector that transforms biased individual estimates into an unbiased collective estimate.
Assumption~\ref{A4} suggests that such decentralized bias correction requires a sufficient level of diversity in the influence network. This condition naturally fails in echo chambers and is consistent with the principle that diversity of perspectives promotes collective intelligence~\cite{Page2007}.

\begin{example}(Performance of dSPRR Algorithm)\label{exa4}
We illustrate the dSPRR Algorithm using the same $W$, $\theta=0$, and $\bd{\delta}=(-2, 5, 3,-6,1)^{\top}$ as in Example~\ref{exa3}. Fig.~\ref{Fig5}(a) shows the local pairing structure over $\mathcal{G}(W)$, where nodes $1$--$5$ select nodes $3$, $4$, $1$, $2$, and $1$,
respectively, to compute the mixing weights. As a result, the pairing structure yields $\bd{x}^{*}=(\frac{23}{75},\ \frac{12}{55},\ \frac{4}{25},\ \frac{2}{11},\ \frac{2}{15})^{\top}$. Fig.~\ref{Fig5}(b) shows the trajectories of $y_{\textup{ave}}(k)$ with $\hat{\bd{a}}^{*}$ generated by Algorithm~\ref{alg:dsspa} under different initializations of $\bd{z}(0)$ and stopping times $T$. We observe that as $T$ increases, the trajectories converge to $0$, while different initializations yield different errors under the same $T$, consistent with 
Proposition~\ref{P5}. 
\begin{figure}
\centering

\begin{minipage}{0.38\linewidth}
    \centering
    \includegraphics[width=\linewidth]{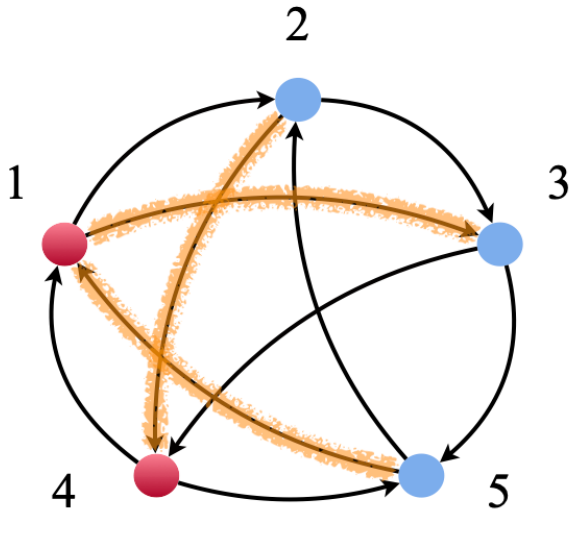}
    
    (a) 
\end{minipage}
\hfill
\begin{minipage}{0.6\linewidth}
    \centering
    \includegraphics[width=\linewidth]{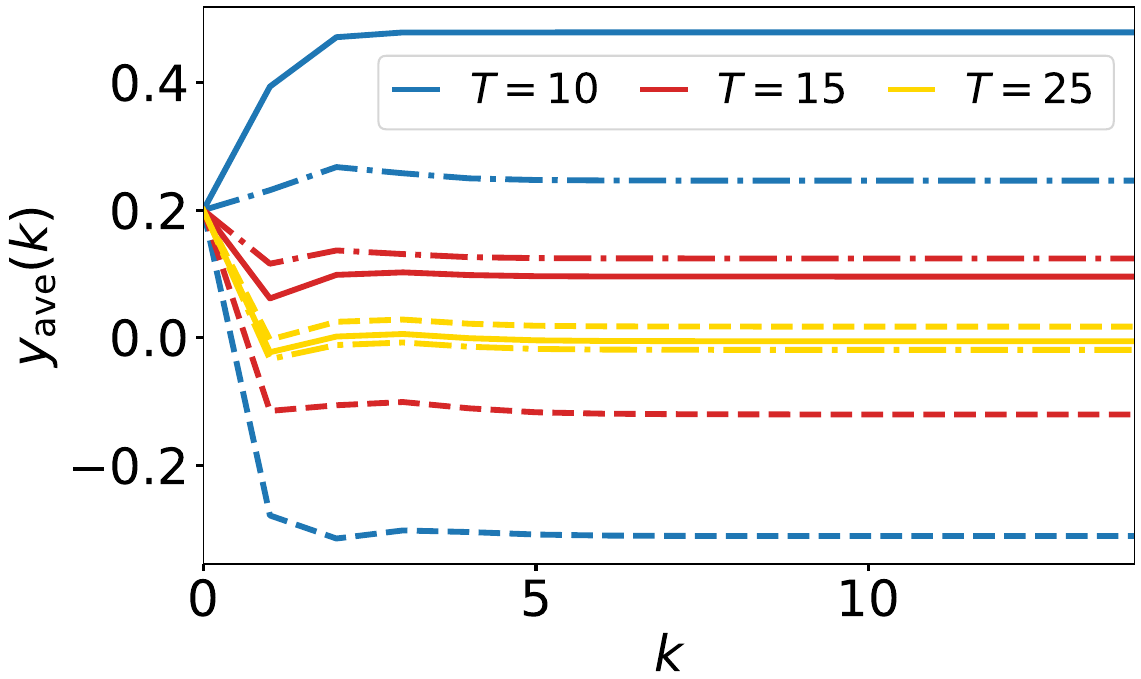}
    
    (b) 
\end{minipage}
\caption{Example~\ref{exa4}. (a) Local pairing structure between agents with opposite biases over $\mathcal{G}(W)$. (b) 
  Trajectories of $y_{\textup{ave}}(k)$ under different initializations and stopping times.} 
\label{Fig5}
\end{figure}

\end{example}

\section{Conclusion}\label{S7}

This paper has investigated the problem of susceptibility optimization for maximizing the wisdom of crowds under the FJ model. To address the resulting nonconvex optimization problem, we proposed the SPRR framework, which consists of social power reformulation and social power
realization. We characterized the optimal social power sets under different bias structures and developed efficient construction algorithms. We further introduced the notions of susceptibility realizability, established necessary and sufficient conditions, and derived closed-form optimal susceptibility vectors. These results provide a unified algorithmic framework with provable complexity guarantees and a fully distributed implementation.

Beyond the specific optimization problem considered in this paper, the proposed SPRR framework establishes a connection between susceptibility and social power, which can be applied to other collective decision-making tasks where the objective is governed by social power. Moreover, the decentralized SPRR provides a mechanism explaining how globally optimal collective outcomes emerge from local interactions.
Future work will investigate susceptibility optimization under time-varying networks and budgeted susceptibilities.

\appendices

\section{}\label{app:L}

\subsection{Proof of Proposition~\ref{Pmono}}
By $\bd{x}=V^{\top}\bd{1}_{n}/n$, we have 
\begin{align*}
  (I_{n}-\dg{\bd{a}})^{-1}\bd{x}-W^{\top}\dg{\bd{a}}(I_{n}-\dg{\bd{a}})^{-1}\bd{x}=\frac{\bd{1}_{n}}{n}.
  \end{align*}
Taking the derivatives with respect to $a_{i}$, we obtain 
\begin{align*}
&(I_{n}-\dg{\bd{a}})^{-1}\frac{\partial \bd{x}}{\partial a_i}
+ \dg{\bd{e}_i}(I_{n}-\dg{\bd{a}})^{-2}\bd{x} \\
&\quad
- W^\top \dg{\bd{a}}(I_{n}-\dg{\bd{a}})^{-1}\frac{\partial \bd{x}}{\partial a_i}
- W^\top \dg{\bd{e}_i}(I_{n}-\dg{\bd{a}})^{-2}\bd{x}= 0,
\end{align*}
which implies
\begin{align}\label{eq7}
  \frac{\partial \bd{x}}{\partial a_{i}}
  =-\frac{x_{i}}{(1-a_{i})^{2}}V^{\top}(I_{n}-W^{\top})\bd{e}_{i}.
\end{align}
Here, $x_{i}>0$ follows from Lemma~\ref{LV}~\ref{LV-5}.

Regarding~\ref{L1-1}, it follows from~\eqref{eq7} that 
\begin{align*}
  \frac{\partial x_{i}}{\partial a_{i}}=\bd{e}^{\top}_{i}\frac{\partial \bd{x}}{\partial a_{i}}=&-\frac{x_{i}}{(1-a_{i})^{2}}\bd{e}^{\top}_{i}V^{\top}(I_{n}-W^{\top})\bd{e}_{i}\\
  =&-\frac{x_{i}}{(1-a_{i})^{2}}(V_{ii}-\sum_{j=1}^{n}W_{ij}V_{ji})< 0,
\end{align*}
where the inequality is implied by $W_{ii}<1$ and Lemma~\ref{LV}~\ref{LV-3}. 

Regarding~\ref{L1-2}, similarly, we have 
\begin{align*}
  \frac{\partial x_{j}}{\partial a_{i}}=\bd{e}^{\top}_{j}\frac{\partial \bd{x}}{\partial a_{i}}=-\frac{x_{i}}{(1-a_{i})^{2}}(V_{ij}-\sum_{l=1}^{n}W_{il}V_{lj}).
\end{align*}
By $(I_{n}-\dg{\bd{a}}W)V=I_{n}-\dg{\bd{a}}$, we obtain $V_{ij}=a_{i}\sum_{l=1}^{n}W_{il}V_{lj}$. As a result, 
\begin{align*}
  \frac{\partial x_{j}}{\partial a_{i}}=-\frac{x_{i}}{(1-a_{i})^{2}}(V_{ij}-\sum_{l=1}^{n}W_{il}V_{lj})=\frac{x_{i}}{a_{i}(1-a_{i})}V_{ij}\geq 0.
\end{align*}
Moreover, Lemma~\ref{LV}~\ref{LV-4} implies that the above inequality is strict if and only if there is a PSP from $i$ to $j$.

\subsection{Proof of Lemma~\ref{L5}}

Since $f(\bd{a})=(\bd{x}^{\top}\bd{\delta})^2$, we have 
\begin{align*}
\frac{\partial f}{\partial \bd{a}}=2(\bd{x}^{\top}\bd{\delta})\left(\frac{\partial \bd{x}}{\partial \bd{a}}\right)^{\top}\bd{\delta},
\end{align*} 
where $\partial \bd{x}/\partial \bd{a}$ denotes the Jacobian of $\bd{x}$ with respect to $\bd{a}$. By~\eqref{eq7}, we obtain 
\begin{align*}
  \frac{\partial \bd{x}}{\partial \bd{a}}=\left[\frac{\partial \bd{x}}{\partial a_{1}}\;\dots\;\frac{\partial \bd{x}}{\partial a_{n}}\right]=-V^{\top}(I_{n}-W^{\top})\dg{\bd{x}}(I_{n}-\dg{\bd{a}})^{-2}.
\end{align*}
Thus,
\begin{align*}
  \frac{\partial f}{\partial \bd{a}}=&-2(\bd{x}^{\top}\bd{\delta})\dg{\bd{x}}(I_{n}-\dg{\bd{a}})^{-2} (I_{n}-W)V\bd{\delta}.
\end{align*}
Moreover, observe that 
\begin{align*}
  &(I_{n}-W)(I_{n}-\dg{\bd{a}}W)^{-1}\\
  =&(I_{n}-\dg{\bd{a}}W+\dg{\bd{a}}W-W)(I_{n}-\dg{\bd{a}}W)^{-1}\\
  =&I_{n}-(I_{n}-\dg{\bd{a}})W(I_{n}-\dg{\bd{a}}W)^{-1},
\end{align*}
which implies $(I_{n}-W)V=(I_{n}-\dg{\bd{a}})(I_{n}-WV)$. Therefore, 
\begin{align*}
  \frac{\partial f}{\partial \bd{a}}=-2(\bd{x}^{\top}\bd{\delta})\dg{\bd{x}}(I_{n}-\dg{\bd{a}})^{-1}(I_{n}-WV)\bd{\delta}.
  \end{align*}
\subsection{Proof of Proposition~\ref{P4}}

\emph{Necessity}.  We show that if Assumption~\ref{A1} holds and $\bd{x}$ is the social power of the FJ model, then~\eqref{eq12} holds for some $\bd{\gamma}\geq 0$. By Lemma~\ref{L0}, $\rho(\dg{\bd{a}}W)<1$ under Assumption~\ref{A1}. Hence, $\bd{\psi}=(I_{n}-W^{\top}\dg{\bd{a}})^{-1}\bd{1}_{n}/n\geq 0$ is well-defined, and $\bd{x}=(I_{n}-\dg{\bd{a}})\bd{\psi}$ follows from~\eqref{equ:sp}. Let $\bd{\gamma}=\dg{\bd{a}}\bd{\psi}$. Then, $\bd{\gamma}\geq 0$ and $\bd{x}+\bd{\gamma}=\bd{\psi}$, which yields  
\begin{align*}
  \dg{\bd{a}}(\bd{x}+\bd{\gamma})=\dg{\bd{a}}\bd{\psi}=\bd{\gamma}.
\end{align*}
Moreover, it follows from $(I_{n}-W^{\top}\dg{\bd{a}})\bd{\psi}=\bd{1}_{n}/n$ that 
\begin{align*}
  L^{\top}\bd{\gamma}&=(I_{n}-W^\top)\dg{\bd{a}}\bd{\psi}\\
  &=(\dg{\bd{a}}-I_{n}+I_{n}-W^{\top}\dg{\bd{a}})\bd{\psi}\\
  &=\frac{\bd{1}_{n}}{n}-\bd{x}.
\end{align*}
Therefore,~\eqref{eq12} holds for $\bd{\gamma}=\dg{\bd{a}}\bd{\psi}$.

\emph{Sufficiency}. We first show that if~\eqref{eq12} holds for some
$\bd{\gamma}\geq 0$, then Assumption~\ref{A1} holds. Let
$\mathcal{S}$ be a sink SCC of $\mathcal{G}(W)$. Since
$W_{ji}=0$ for all $j\in\mathcal{S}$ and $i\notin\mathcal{S}$, we
have $\sum_{i\in\mathcal{S}}W_{ji}=1$ for all $j\in\mathcal{S}$.
Then, by~\eqref{eq12b}, we obtain
\begin{align*}
  \sum_{i\in\mathcal{S}}x_i
  &=\frac{|\mathcal{S}|}{n}-\sum_{i\in\mathcal{S}}\gamma_i
  +\sum_{j\in\V}\gamma_j\sum_{i\in\mathcal{S}}W_{ji}\\
  &=\frac{|\mathcal{S}|}{n}+\sum_{j\in\V\setminus\mathcal{S}}
  \gamma_j\sum_{i\in\mathcal{S}}W_{ji}>0,
\end{align*}
where the inequality follows from $\bd{\gamma}\geq 0$ and
$|\mathcal{S}|\geq 1$. Hence, $x_l>0$ for some $l\in\mathcal{S}$. Moreover, by~\eqref{eq12a}, $a_i=1$ only if $x_i=0$, which implies $a_l<1$. Therefore, any sink SCC contains an stubborn agent, which means Assumption~\ref{A1} holds.

Next, we show that $\bd{x}$ is the social power given by~\eqref{equ:sp}.  Since Assumption~\ref{A1} holds,
$(I_{n}-W^{\top}\dg{\bd{a}})^{-1}$ exists. 
By~\eqref{eq12a}, $\bd{x}=(I_{n}-\dg{\bd{a}})(\bd{x}+\bd{\gamma})$. Then, it follows from~\eqref{eq12b} that
\begin{align*}
  (I_{n}-W^{\top}\dg{\bd{a}})(\bd{x}+\bd{\gamma})&=(I_{n}-\dg{\bd{a}}+L^{\top}\dg{\bd{a}})(\bd{x}+\bd{\gamma})\\
  &=\bd{x}+L^{\top}\bd{\gamma}=\frac{\bd{1}_{n}}{n},
\end{align*}
which implies $\bd{x}+\bd{\gamma}=(I_{n}-W^{\top}\dg{\bd{a}})^{-1}\bd{1}_{n}/n$. Therefore,
$\bd{x}=(I_{n}-\dg{\bd{a}})(I_{n}-W^{\top}\dg{\bd{a}})^{-1}\bd{1}_{n}/n$, which is exactly the social power of the
FJ model.

\section{}\label{app2}
\subsection{Proof of Theorem~\ref{T0}}
Let $\mathcal{D}_{\bd{u}}=\setdef{\bd{z}\in \mathbb{R}^n}{0\leq\bd{z}\leq\bd{u}}$. For $\bd{a}\in\mathcal{D}_{\bd{u}}$, since Assumption~\ref{A1} holds, we have $(I_{n}-\dg{\bd{a}}W)V=I_{n}-\dg{\bd{a}}$, and thus $\dg{\bd{a}}(I_{n}-WV)=I_{n}-V$. Hence, whenever $a_i>0$, by continuity and Lemma~\ref{L5}, we obtain 
\begin{align}\label{equ13}
\frac{\partial f}{\partial a_{i}}=2(\bd{x}^{\top}\bd{\delta})\frac{x_i}{a_i(1-a_i)}(\sum_{j=1}^{n}V_{ij}\delta_{j}-\delta_{i}),
\end{align}
where $\bd{x}^{\top}\bd{\delta}>0$, and $x_i>0$ follows from Lemma~\ref{LV}~\ref{LV-5}.

Regarding~\ref{P3-1}, suppose that $\bd{a}^{*}$ is optimal 
with $a^{*}_i>0$ for some $i\in\mathcal{J}$, and let $j\notin\mathcal{I}_{\min}^{\bd{\delta}}$ satisfy $W_{ij}>0$. Since
the edge $(i,j)$ is a PSP, Lemma~\ref{LV}~\ref{LV-4} yields
$V_{ij}>0$ with $\delta_j>\delta_i$. By~\eqref{equ13},
$\partial f/\partial a_{i}>0$ at $\bd{a}^{*}$, which contradicts the optimality of $\bd{a}^{*}$. Therefore, $a^{*}_i=0$ for all $i\in\mathcal{J}$.

Regarding~\ref{P3-2}, we first prove the sufficiency. Since $V$ is row-stochastic, by~\eqref{equ13}, $\partial f/\partial a_{i}\geq 0$ for all $i\in\mathcal{I}_{\min}^{\bd{\delta}}$, which means $f$ is nondecreasing in $a_i$ for $i\in\mathcal{I}_{\min}^{\bd{\delta}}$. 
Thus, there exists a global minimum on the face 
$\mathcal{F}_{\mathcal{D}_{\bd{u}}}=\setdef{\bd{z}\in \mathcal{D}_{\bd{u}}}{z_{i}=0, \; \forall i\in\mathcal{I}_{\min}^{\bd{\delta}}}$. Next, we focus on $\min_{\bd{a}\in\mathcal{F}_{\mathcal{D}_{\bd{u}}}} f(\bd{a})$. 
Since 
\begin{align*}
  I_{n}-V=&(I_{n}-\dg{\bd{a}}W)^{-1}(I_{n}-\dg{\bd{a}}W-I_{n}+\dg{\bd{a}})\\=&(I_{n}-\dg{\bd{a}}W)^{-1}\dg{\bd{a}}L,
\end{align*} 
we have 
\begin{align*}
  I_{n}-WV=&I_{n}-W+W-WV=L+W(I_{n}-V)\\
  =&(I_{n}+W(I_{n}-\dg{\bd{a}}W)^{-1}\dg{\bd{a}})L.
\end{align*}
Let $\Psi=W(I_{n}-\dg{\bd{a}}W)^{-1}\dg{\bd{a}}$. By Lemma~\ref{L0}, on $\mathcal{F}_{\mathcal{D}_{\bd{u}}}$, $\Psi$ is a nonnegative matrix whose columns indexed by $\mathcal{I}_{\min}^{\bd{\delta}}$ are zero. Moreover, by continuity, Lemma~\ref{L5} implies
\begin{align*}
\frac{\partial f}{\partial \bd{a}}=-2(\bd{x}^{\top}\bd{\delta})\dg{\bd{x}}(I_{n}-\dg{\bd{a}})^{-1}(I_{n}+\Psi)\bd{\xi} 
\end{align*} 
with $\bd{\xi}=L\bd{\delta}$. Therefore, for all $i\notin\mathcal{I}_{\min}^{\bd{\delta}}$, we obtain 
\begin{align}\label{equ14}
\frac{\partial f}{\partial a_{i}}=-2(\bd{x}^{\top}\bd{\delta})\frac{x_i}{1-a_i}(\xi_i+\sum_{j\notin\mathcal{I}_{\min}^{\bd{\delta}}}\Psi_{ij}\xi_{j})\leq 0,
\end{align} 
where the inequality only requires $\xi_{i}\geq 0$. Consequently, $\bar{\bd{a}}\in\mathcal{F}_{\mathcal{D}_{\bd{u}}}$ with $\bar{a}_i=u_i$ for all $i\notin\mathcal{I}_{\min}^{\bd{\delta}}$ is an optimal solution. 
Note that for $\bd{a}^*$ satisfying~\eqref{opti}, and for any $i\in\mathcal{I}_{\min}^{\bd{\delta}}\setminus\mathcal{J}$ and $j\notin\mathcal{I}_{\min}^{\bd{\delta}}$,
every path from $i$ to $j$ passes through some agent $l\in\mathcal{J}$ with $a_l^*=0$. Thus, Lemma~\ref{LV}~\ref{LV-4} yields $V_{ij}=0$ whenever $a^*_i>0$, which, by~\eqref{equ13}, implies $\partial f/\partial a_{i}= 0$. Therefore, by continuity, $f$ is constant over all $\bd{a}^*$ satisfying~\eqref{opti}. Since $\bar{\bd{a}}$ satisfies~\eqref{opti}, any $\bd{a}^*$ satisfying~\eqref{opti} is also optimal. 

For the necessity, by~\ref{P3-1}, all optimal solutions
lie in $\mathcal{F}_{\mathcal{J}}=\setdef{\bd{z}\in
\mathcal{D}_{\bd{u}}}{z_{j}=0, \; \forall j\in\mathcal{J}}$, on which $\Psi_{ij}=0$ whenever $j\in\mathcal{J}$. Moreover, $\xi_j=0$ for all $j\in\mathcal{I}_{\min}^{\bd{\delta}}\setminus\mathcal{J}$, since $W_{jl}=0$ for all $l\notin\mathcal{I}_{\min}^{\bd{\delta}}$. Hence, by~\eqref{equ12}, \eqref{equ14} strictly holds on $\mathcal{F}_{\mathcal{J}}$ for all $i\notin\mathcal{I}_{\min}^{\bd{\delta}}$. That is, $f$ is strictly decreasing in $a_i$ for all $i\notin\mathcal{I}_{\min}^{\bd{\delta}}$. Therefore, every optimal
solution $\bd{a}^*$ satisfies $\bd{a}^*\in\mathcal{F}_{\mathcal{J}}$ and $a^{*}_i=u_i$ for all $i\notin\mathcal{I}_{\min}^{\bd{\delta}}$, and thus~\eqref{opti}. 
    
\subsection{Proof of Theorem~\ref{T1}}
Regarding~\ref{T1-1}, for $\nu\in\mathbb{R}$ and $\bd{\lambda}\in\mathbb{R}^n$, let 
\begin{align*}
  \mathcal{L}(\bd{x},\nu,\bd{\lambda})=\bd{x}^{\top}\bd{\delta}\bd{\delta}^{\top}\bd{x}+\nu (\bd{x}^{\top}\bd{1}_{n}-1)-\bd{\lambda}^{\top}\bd{x}.
\end{align*}
By the KKT conditions, any optimal solution $\bd{x}^{*}$ satisfies
\begin{align*}
2(\bd{\delta}^{\top}\bd{x}^{*})\bd{\delta}+\nu\bd{1}_n=\bd{\lambda}
\end{align*}
for some $\nu$ and $\bd{\lambda}\geq 0$ with $\bd{\lambda}^{\top}\bd{x}^{*}=0$. Suppose that $c=\bd{\delta}^{\top}\bd{x}^{*}\neq 0$. Multiplying both sides by
$\bd{x}^{*\top}$ yields $\nu=-2c^{2}$, and thus
$c(\delta_i-c)=\lambda_i/2\geq 0$ for all $i\in\V$, which contradicts $\min_{i}\delta_{i}<0<\max_{i}\delta_{i}$. Therefore, $\bd{\delta}^{\top}\bd{x}^{*}=0$, and the optimal solution set is
$\mathcal{X}=\setdef{\bd{x}\in\Delta_n}{\bd{x}^{\top}\bd{\delta}=0}$.
 
Regarding~\ref{T1-2}, by $\min_{i}\delta_{i}<0<\max_{i}\delta_{i}$, for any $i\in\V$, there exists a $q(i)\in\V$ such that $\delta_{i}\delta_{q(i)}<0$. Let 
\begin{align*}
 \eta_{i}=\frac{\delta_{q(i)}}{\delta_{q(i)}-\delta_{i}}. 
\end{align*}
Then, we have $0<\eta_{i}<1$, $\eta_{i}\bd{e}_{i}+(1-\eta_{i})\bd{e}_{q(i)}\in\Delta_n$, and 
\begin{align*}
 \bd{\delta}^{\top}(\eta_{i}\bd{e}_{i}+(1-\eta_{i})\bd{e}_{q(i)})=0,
\end{align*}
which means $\eta_{i}\bd{e}_{i}+(1-\eta_{i})\bd{e}_{q(i)}\in\mathcal{X}$. Since $\mathcal{X}$ is convex, for any $\bd{z}\in\interior\Delta_n$, we obtain 
\begin{align*}
 \bd{x}^{*}=\sum_{i=1}^{n}z_{i}(\eta_{i}\bd{e}_{i}+(1-\eta_{i})\bd{e}_{q(i)})\in\mathcal{X}.
 \end{align*}
Moreover, since $0<\eta_{i}<1$ for all $i\in\V$, $\bd{x}^{*}\in\interior\Delta_n$ follows from $\bd{z}\in\interior\Delta_n$.

It remains to prove the running time. All steps of the algorithm except Step $2$ clearly require a single pass over the nodes and therefore take $\mathcal{O}(n)$ time. Let $n^{+}=|\mathcal{V}^{+}|$, $n^{-}=|\mathcal{V}^{-}|$, and assume $n^{+}>n^{-}$. Since $\mathcal{G}$ is a complete bipartite graph, a (minimum) edge cover can be obtained by pairing each node in $\mathcal{V}^{-}$ with a distinct node in $\mathcal{V}^{+}$ and then pairing each of the remaining $n^{+}-n^{-}$ nodes in $\mathcal{V}^{+}$ with an arbitrary node in $\mathcal{V}^{-}$. This construction requires $\mathcal{O}(n)$ time. Hence, the algorithm runs in $\mathcal{O}(n)$ time.
\section{Proof of Theorem~\ref{T3}}\label{app:T3}
Regarding~\ref{T3-1}, by Proposition~\ref{P4}, the FJ model is FSR if and only if for any $\bd{x}\in\Delta_{n}$, 
there exists $\bd{\gamma}\geq 0$ such that 
\begin{align}\label{eq19}
  L^{\top}\bd{\gamma}=\bd{b}
\end{align}
with $\bd{b}=\bd{1}_{n}/n-\bd{x}$. 
Note that~\eqref{eq19} has at least one solution if and only if   
\begin{align*}
  \bd{b}\in\mathbf{Range}(L^{\top})=\mathbf{Null}(L)^{\perp }.
\end{align*} 
That is, $\bd{b}^{\top}\bd{z}=0$ for any $\bd{z}\in\mathbf{Null}(L)$, which holds for any $\bd{x}\in\Delta_{n}$ if and only if $\mathbf{Null}(L)=\Span\{\bd{1}_{n}\}$. Hence,~\eqref{eq19} has at least one solution for any $\bd{x}\in\Delta_{n}$ if and only if $\mathcal{G}(W)$ has a unique sink SCC. Assume that $\mathcal{G}(W)$ has a unique sink SCC. By~\cite[Theorem 2.1.2]{SLC-CDM:09}, all solutions of~\eqref{eq19} can be written as
\begin{align*}
  \bd{\gamma}=(L^{\top})^{\dagger}\bd{b}+\bd{\gamma}^{\prime}
\end{align*}
with $\bd{\gamma}^{\prime}\in\mathbf{Null}(L^{\top})$. 

If $\mathcal{G}(W)$ is strongly connected, we have $\mathbf{Null}(L^{\top})=\Span\{\bd{\omega}\}$, where $\bd{\omega}>0$ is the dominant left eigenvector of $W$ associated with eigenvalue $1$. Therefore, for any $\bd{b}\in\mathbb{R}^{n}$, there exists a constant $\beta\in\mathbb{R}$ such that 
\begin{align}\label{eq20}
  \bd{\gamma}^{*}=(L^{\top})^{\dagger}\bd{b}+\beta\bd{\omega}\geq 0.
\end{align}
If $\mathcal{G}(W)$ is not strongly connected, $W$ and $L$ can be written in the following block form  
\begin{align*}
W=\begin{bmatrix}
   W_{1} & 0 \\
W_{3} & W_{2}
\end{bmatrix}, \quad L=\begin{bmatrix}
   L_{1} & 0 \\
L_{3} & L_{2}
\end{bmatrix},
\end{align*}
where $L_{1}\in\mathbb{R}^{m\times m}$ corresponds to the unique sink SCC with dominant left eigenvector $\bd{\omega}^{1}\in\interior\Delta_{m}$ satisfying $L_{1}^{\top}\bd{\omega}^{1}=0$, and $L_{2}=I_{n-m}-W_{2}$ with $\rho(W_{2})<1$. Let  
\begin{align*}
  \bd{\omega}^{\prime}=\begin{bmatrix}
   \bd{\omega}^{1}\\
\bd{0}_{n-m}
\end{bmatrix}.
\end{align*}
Then, $\bd{\omega}^{\prime}\in \mathbf{Null}(L^{\top})$.
On the other hand, since $\mathcal{G}(W)$ has a unique sink SCC, the dimension of $\mathbf{Null}(L^{\top})$ is $1$~\cite[Lemma 2]{YT-LW:23}. Hence, $\mathbf{Null}(L^{\top})=\Span\{\bd{\omega}^{\prime}\}$. Moreover, let
\begin{align*}
G=\begin{bmatrix}
   L_{1}^{\dagger} & 0 \\
-L_{2}^{-1}L_{3}L_{1}^{\dagger} & L_{2}^{-1}
\end{bmatrix},
\end{align*}
which satisfies $LGL=L$, and thus $L^{\dagger}=L^{\dagger}LGLL^{\dagger}$. Let $\bd{u}=\bd{1}_n/\sqrt{n}$ and
$\bd{v}=\bd{\omega}^{\prime}/\|\bd{\omega}^{\prime}\|$. Then, $L^{\dagger}L=I_n-\bd{u}\bd{u}^{\top}$ and
$LL^{\dagger}=I_n-\bd{v}\bd{v}^{\top}$. Therefore, for
$\bd{b}=[\bd{b}_1^\top \ \bd{b}_2^\top]^\top$ with
$\bd{b}_{1}\in\mathbb{R}^{m}$, 
\begin{align*}
(L^{\top})^{\dagger}\bd{b}&=(I_n-\bd{v}\bd{v}^{\top})G^{\top}
(I_n-\bd{u}\bd{u}^{\top})\bd{b}\\
&=\begin{bmatrix}
   \times & \times \\
0 & (L_{2}^{-1})^\top
\end{bmatrix}\begin{bmatrix}
   \bd{b}_1  \\
\bd{b}_2
\end{bmatrix},
\end{align*}
where $\times$ denotes blocks irrelevant to the analysis, and the last equality follows from $\bd{u}^{\top}\bd{b}=0$ and $v_i=0$ for all $i>m$. Thus, the last $n-m$ entries of $(L^{\top})^{\dagger}\bd{b}$ are $(L_{2}^{-1})^{\top}\bd{b}_{2}$, with $L_{2}^{-1}=\sum_{t=0}^{\infty}W_{2}^{t}\geq 0$.
Consequently, for any $\bd{x}\in\Delta_{n}$ with $x_i>1/n$ for all $i>m$, we have $\bd{b}_{2}<0$ and thus
$(L_{2}^{-1})^{\top}\bd{b}_{2}<0$. Since
$\omega_{i}^{\prime}=0$ for all $i> m$, there does not exist
$\beta\in\mathbb{R}$ such that
$(L^{\top})^{\dagger}\bd{b}+\beta\bd{\omega}^{\prime}\geq 0$. Hence,
\eqref{eq19} has a nonnegative solution for any $\bd{x}\in\Delta_{n}$
if and only if $\mathcal{G}(W)$ is strongly connected.

Regarding~\ref{T3-2}, by~\ref{T3-1}, every
$\bd{x}^{*}\in\mathcal{X}$ can be induced, and thus
$\bd{a}^{*}$ solves problem~\eqref{eq3} if and only if it induces some
$\bd{x}^{*}\in\mathcal{X}$. By~\eqref{eq20}, this holds if and only if
$\bd{\gamma}^{*}=(L^{\top})^{\dagger}\bd{b}^{*}+\beta\bd{\omega}$ for
some $\beta$ ensuring $\bd{\gamma}^{*}\geq 0$, and
$\bd{a}^{*}=\dg{\bd{x}^{*}+\bd{\gamma}^{*}}^{-1}\bd{\gamma}^{*}$ by~\eqref{eq12a}.

\section{}\label{app:T4}
\subsection{Proof of Theorem~\ref{T4}}
Let $P=\bd{\omega} \bd{1}_{n}^{\top}$ and $U=M_{\alpha}-P$. Then, we have 
\begin{align*}
  &M_{\alpha}\bd{\omega}=\bd{\omega}, \quad \quad \bd{1}_{n}^{\top}M_{\alpha}=\bd{1}_{n}^{\top}, \\
  &P\bd{b}^{*}=0, \quad \quad UP=PU=0.
\end{align*}

Regarding~\ref{T4-1}, since $W$ is irreducible, $M_{\alpha}$ is column-stochastic and primitive. Hence, 
\begin{align*}
  \lim_{k\to\infty}\tilde{\bd{\omega}}(k)=\lim_{k\to\infty}M_{\alpha}^{k}\tilde{\bd{\omega}}(0)
  =P\tilde{\bd{\omega}}(0)=\bd{\omega} \bd{1}_{n}^{\top}\tilde{\bd{\omega}}(0).
\end{align*} 
Thus, $\lim_{k\to\infty}\tilde{\bd{\omega}}(k)=\bd{\omega}$ if and only if $\tilde{\bd{\omega}}^{\top}(0)\bd{1}_{n}=1$.

Regarding~\ref{T4-2}, for system~\eqref{eqz}, we have 
\begin{align*}
\bd{z}(k)&=M_{\alpha}^{k}\bd{z}(0)+\alpha\sum_{t=0}^{k-1}M_{\alpha}^{t}\bd{b}^{*}\\
&=(U+P)^{k}\bd{z}(0)+\alpha\sum_{t=0}^{k-1}(U+P)^{t}\bd{b}^{*},\\
&=P\bd{z}(0)+U^{k}\bd{z}(0)+\alpha\sum_{t=0}^{k-1}U^{t}\bd{b}^{*},
\end{align*}
where the last equality follows from $UP=PU=0$, $P\bd{b}^{*}=0$, $P^{2}=P$, and $(U+P)^{t}=U^{t}+P$ for all $t>0$. Next, we show that $\rho(U)<1$. As $M_{\alpha}$ is column-stochastic and primitive, let $1,\lambda_2,\dots,\lambda_n$ be its eigenvalues, where $|\lambda_i|<1$ for all $i\geq 2$. It follows from $M_{\alpha}\bd{\omega}=\bd{\omega}$ that 
\begin{align*}
 \bd{\omega}=(U+P)\bd{\omega}=U\bd{\omega}+\bd{\omega},
\end{align*}
which yields $U\bd{\omega}=0$. That is, 
$0$ is an eigenvalue of $U$. For $\lambda_i\neq 1$ with eigenvector $\bd{\pi}$, by $\bd{1}_{n}^{\top}M_{\alpha}=\bd{1}_{n}^{\top}$, we obtain 
\begin{align*}
\bd{1}_{n}^{\top} \bd{\pi}= \bd{1}_{n}^{\top}M_{\alpha}\bd{\pi}=\lambda_{i}\bd{1}_{n}^{\top}\bd{\pi},
\end{align*}
which implies $\bd{1}_{n}^{\top} \bd{\pi}=0$. As a result, $P\bd{\pi}=0$ and 
\begin{align*}
 U \bd{\pi}=(M_{\alpha}-P)\bd{\pi}=M_{\alpha}\bd{\pi}=\lambda_{i}\bd{\pi},
\end{align*}
which means $\lambda_i$ is an eigenvalue of $U$. Thus, the eigenvalues of $U$ are $0,\lambda_2,\dots,\lambda_n$, which implies $\rho(U)<1$. Hence,  
\begin{align}\label{eq23}
  \bd{z}^{*}=\lim_{k\to\infty}\bd{z}(k)=&P\bd{z}(0)+\lim_{k\to\infty}\left(U^{k}\bd{z}(0)+\alpha\sum_{t=0}^{k-1}U^{t}\bd{b}^{*}\right)\notag\\
  =&P\bd{z}(0)+\alpha (I_{n}-U)^{-1}\bd{b}^{*}.
\end{align}
Moreover, by $L^{\top}\bd{\omega}=0$, $\rho(U)<1$, and $PU=0$, we have 
\begin{align*}
  L^{\top}\bd{z}^{*}=&L^{\top}P\bd{z}(0)+\alpha L^{\top}(I_{n}-U)^{-1}\bd{b}^{*}\\
  =&\lim_{k\to\infty}(I_{n}-M_{\alpha})\sum_{t=0}^{k-1}U^{t}\bd{b}^{*}\\=&\lim_{k\to\infty}(I_{n}-U-P)\sum_{t=0}^{k-1}U^{t}\bd{b}^{*}.
\end{align*}
Since $PU^{t}\bd{b}^{*}=0$ for all $t\geq 0$, we obtain 
\begin{align*}
  L^{\top}\bd{z}^{*}=&\lim_{k\to\infty}(I_{n}-U)\sum_{t=0}^{k-1}U^{t}\bd{b}^{*}\\
  =&\bd{b}^{*}-\lim_{k\to\infty}U^{k}\bd{b}^{*}=\bd{b}^{*},
\end{align*}
where the last equality is implied by $\rho(U)<1$. 
Thus, $\bd{z}^{*}$ is a solution of $L^{\top}\bd{z}=\bd{b}^{*}$. 

Regarding~\ref{T4-3} and~\ref{T4-4}, since $\bd{z}^{*}$ is a solution of $L^{\top}\bd{z}=\bd{b}^{*}$, all its solutions can be written as $\bd{z}=\bd{z}^{*}+\beta \bd{\omega}$. 
Hence, $\bd{z}\geq 0$ for any $\beta\geq \max_{i}-z_{i}^{*}/\omega_{i}$. By Theorem~\ref{T3}, this yields a solution of problem~\eqref{eq3}. Moreover, by~\eqref{eq23}, $\bd{z}^{*}\geq 0$ if and only if $\bd{\omega}\bd{1}_{n}^{\top}\bd{z}(0)\geq -\alpha \bd{\phi}$, i.e., $\bd{1}_{n}^{\top}\bd{z}(0)\geq \max_{i}-\alpha\phi_{i}/\omega_{i}$. In this case, $\beta=0$ is admissible, and thus
$\bd{a}^{*}=\dg{\bd{x}^{*}+\bd{z}^{*}}^{-1}\bd{z}^{*}$ solves problem~\eqref{eq3}.

\subsection{Proof of Proposition~\ref{P5}}
Regarding~\ref{P5-1}, at each iteration, agents exchange information only with their neighbors, resulting in $\mathcal{O}(|\mathcal{E}|)$ communications per iteration. Hence,  the total communication cost is $\mathcal{O}(T|\mathcal{E}|)$ over $T$ iterations.

Regarding~\ref{P5-2}, the max-consensus algorithm in Steps~$7,8$ propagates the global maximum by at least one hop per iteration~\cite{Cortes2008Automatica}, and therefore converges in at most $\textup{diam}(\mathcal{G}(W))$ iterations, where $\textup{diam}(\mathcal{G}(W))$ is the diameter of $\mathcal{G}(W)$. Hence, $\bd{\chi }= \bd{z}(T)+\beta_{T}\tilde{\bd{\omega}}(T)$ with $\beta_{T}=\max_{i}(-z_{i}(T)/\tilde{\omega}_{i}(T))$ for all $T\geq \textup{diam}(\mathcal{G}(W))$. Moreover, by Theorem~\ref{T4}, there exists a constant $C_{1}$ such that
\begin{align*}
  \|\bd{z}(T)-\bd{z}^*\|_2 &\leq C_{1}\rho_{2}^{T}(M_\alpha)\|\bd{z}(0)-\bd{z}^*\|_2,\\
  \|\tilde{\bd{\omega}}(T)-\bd{\omega}\|_2&\leq C_{1}\rho_{2}^{T}(M_\alpha)\|\bd{1}_{n}/n-\bd{\omega}\|_2\leq C_{1}\rho_{2}^{T}(M_\alpha),
\end{align*}
where $\rho_{2}(M_\alpha)\in (0,1)$ is any number exceeding the second-largest eigenvalue modulus of $M_\alpha$. 
Since $\beta_{T}\to\beta^{*}$ as $T\to\infty$, there exists $\hat{T}>0$ for any $\hat{\epsilon}\in (0,\epsilon)$, such that $|\beta^{*}-\beta_{T}|\leq x^{*}_{\min}\hat{\epsilon}$ for all $T\geq \hat{T}$, where $x^{*}_{\min}=\min_{i}x^{*}_{i}$.
Let $\bd{\chi }^{*}=\bd{z}^{*}+\beta^{*}\bd{\omega}$. Then, 
\begin{align*}
 &\norm{\bd{\chi }-\bd{\chi }^{*}}_{2}\\\leq &\|\bd{z}(T)-\bd{z}^*\|_2+|\beta_{T}-\beta^{*}|\|\tilde{\bd{\omega}}(T)\|_2+|\beta^{*}|\|\tilde{\bd{\omega}}(T)-\bd{\omega}\|_2 \\
 \leq & C_{1}(\|\bd{z}(0)-\bd{z}^*\|_2+|\beta^{*}|)\rho_{2}^{T}(M_\alpha)+x^{*}_{\min}\hat{\epsilon}.
\end{align*}
Let $C_2=C_{1}(\|\bd{z}(0)-\bd{z}^*\|_2+|\beta^{*}|)/x^{*}_{\min}$. Since 
\begin{align*}
  (\hat{a}_{i}^{*}-a_{i}^{*})^{2}=(\frac{\chi_{i}}{x_{i}^{*}+\chi_{i}}-\frac{\chi_{i}^{*}}{x_{i}^{*}+\chi_{i}^{*}})^{2}\leq \frac{(\chi_{i}-\chi_{i}^{*})^{2}}{(x_{i}^{*})^{2}}, \; \forall i\in\V,
\end{align*}
we have  
\begin{align*}
  \norm{\hat{\bd{a}}^{*}-\bd{a}^{*}}_{2}\leq \frac{1}{x^{*}_{\min}} \norm{\bd{\chi }-\bd{\chi }^{*}}_{2}\leq C_{2}\rho_{2}^{T}(M_\alpha)+\hat{\epsilon}. 
\end{align*}
Thus, $\norm{\hat{\bd{a}}^{*}-\bd{a}^{*}}_{2}\leq \epsilon$ for all
 \begin{align*}
  T\geq\max\left\{\frac{\log\frac{\epsilon-\hat{\epsilon}}{C_2}}{\log(\rho_{2}(M_\alpha))},\ 
\textup{diam}(\mathcal{G}(W)), \hat{T}\right\}.
\end{align*}
\section*{References}
\bibliographystyle{plainurl}
\bibliography{Ye}

\end{document}